\documentclass{aa}  

\usepackage{graphicx}
\usepackage[varg]{txfonts}

\usepackage{lscape}

\defcitealias{clem}{CLHW11}

\newcommand{\nperact}{113}

\begin{document} 
   \title{A detailed understanding of the rotation-activity relationship using the 300\,Myr old open cluster NGC\,3532
                        \thanks{Based on data acquired through the Australian Astronomical Observatory, under program S/2017A/02.}
                        \thanks{Based on observations at Cerro Tololo Inter-American Observatory, National Optical Astronomy Observatory, under proposal 2008A-0476.}
                         \thanks{The full Table 1 is only available in electronic form
                                at the CDS via anonymous ftp to cdsarc.u-strasbg.fr (130.79.128.5)
                                or via http://cdsweb.u-strasbg.fr/cgi-bin/qcat?J/A+A/}
                 }
        
   \author{D. J. Fritzewski\inst{1}
          \and
          S. A. Barnes\inst{1,2}
          \and
          D. J. James\inst{3,4}
          \and
          S. P. J\"arvinen\inst{1}
          \and
          K. G. Strassmeier\inst{1}
          }

   \institute{Leibniz-Institut f\"ur Astrophysik Potsdam (AIP),
              An der Sternwarte 16, 14482 Potsdam, Germany\\
              \email{dfritzewski@aip.de}
         \and
                 Space Science Institute, 4750 Walnut St., Boulder, CO 80301, USA
                \and
                        Center for Astrophysics $\vert$ Harvard \& Smithsonian, 60 Garden Street, Cambridge, MA 02138, USA
                \and
                        Black Hole Initiative at Harvard University, 20 Garden Street, Cambridge, MA 02138, USA
                }
   \date{}

 
  \abstract
   {The coeval stars of young open clusters provide insights into the formation of the rotation-activity relationship that elude studies of multi-age field populations.}
   {We measure the chromospheric activity of cool stars in the 300\,Myr old open cluster NGC~3532 in concert with their rotation periods to study the mass-dependent morphology of activity for this transitional coeval population.}
   {Using multi-object spectra of the \ion{Ca}{ii} infrared triplet region obtained with the AAOmega spectrograph at the 4\,m Anglo-Australian Telescope, we measure the chromospheric emission ratios $R'_\mathrm{IRT}$ for 454 FGKM cluster members of NGC\,3532.
   }
    {The morphology of activity against colour appears to be a near-mirror image of the cluster's rotational behaviour.
        In particular, we identify a group of `desaturated transitional rotators' that branches off from the main group of unsaturated FGK slow rotators, and from which it is separated by an `activity gap'. 
    The few desaturated gap stars are identical to the ones in the rotational gap.
    Nevertheless, the rotation-activity diagram is completely normal. In fact, the relationship is so tight that it allows us to predict rotation periods for many additional stars.
    We then precisely determine these periods from our photometric light curves, allowing us to construct an enhanced colour-period diagram that represents 66\,\% of the members in our sample.
    Our activity measurements show that all fast rotators of near-solar mass (F-G\,type) have evolved to become slow rotators, demonstrating that the absence of fast rotators in a colour-period diagram is not a detection issue but an astrophysical fact. 
    We also identify a new population of low-activity stars among the early M\,dwarfs, enabling us to populate the extended slow rotator sequence in the colour-period diagram.
}
   {The joint analysis of chromospheric activity and photometric time series data thus enables comprehensive insights into the evolution of the rotation and activity of stars
   during the transitional phase between the Pleiades and Hyades ages.}


   \titlerunning{Understanding the rotation-activity relationship in NGC\,3532}
   \authorrunning{D.~J. Fritzewski et al.}

   \maketitle
%
%

\section{Introduction}
The magnetic dynamo of cool stars is believed to be driven jointly by stellar rotation and convection.
Although the magnetic field plays a vital role 
in the creation of various stellar activity phenomena, it is observationally
less accessible than these phenomena.
Because of this difficulty, magnetic activity is typically accessed, especially in larger samples, through certain indicators, such as `excessive' chromospheric emission.
Such emission was first observed by \cite{1913ApJ....38..292E} in the reversal of the \ion{Ca}{II}~H\,and\,K lines in objective-prism spectra of bright stars and was immediately connected to sunspot-like features on other stars.

For the majority of stars, one observes decreasing activity with decreasing rotation rate. 
This correlation between the chromospheric emission and stellar rotation rate can be traced back to \cite{1967ApJ...150..551K} and \cite{1972ApJ...171..565S} and similar studies of the time evolution of different activity tracers. 
\cite{1984ApJ...279..763N} made an important advance, showing that the chromospheric activity of a sample of field stars is particularly well correlated with a mass-normalised version of the rotation period called the Rossby number, 
$\mathrm{Ro} = P_\mathrm{rot}/\tau_c $ 
(with $P_\mathrm{rot}$ representing the stellar rotation period and $\tau_c$ the convective turnover timescale). 
The corresponding reduction in the dimensionality of the problem\footnote{We mean plotting activity vs. Ro, rather than activity vs. both $P_\mathrm{rot}$ and mass or a suitable mass proxy.} allows all cool stars to be considered as part of the same continuum despite differences in their evolutionary timescales.
Indeed, this rotation-activity correlation has repeatedly been shown to be fundamental in understanding stellar activity. 
It is observed not only in chromospheric activity, but also in the coronal X-ray activity of solar-like \citep{2003A&A...397..147P} and fully convective stars \citep{2016Natur.535..526W,2017ApJ...834...85N}.

However, the dimensional reduction at the heart of the relationship also conceals some of the underlying activity behaviours of cool stars.
Certain stars could potentially be over- or under-active as compared with otherwise similar stars.
Likewise, a small change in a single underlying parameter, such as mass or rotation period, could result in a substantial change in activity. 
Such dependences are best addressed in open clusters, where the coevality and homogeneity of the underlying population tame the parameter dependences, permitting more detailed examination than with field star samples.

A number of tracers are available for probing stellar activity in different layers of outer stellar atmospheres. In the photosphere, one observes starspots
(e.g. \citealt{2009A&ARv..17..251S, 2018ApJ...863..190B}) and flares (e.g. \citealt{1989SoPh..121..299P}). 
In the chromosphere, emission in certain spectral lines (e.g. H$\alpha$, \ion{Ca}{ii}~H\,and\,K, and the even bluer \ion{Mg}{ii}~h~and~k) are tracers of heating beyond radiative equilibrium \citep{2008LRSP....5....2H}. 
Finally, in the corona, the outer layer of the stellar atmosphere, emission in soft X-rays is the most prominent activity tracer
\citep{1997A&A...318..215S, 2004A&ARv..12...71G}.

The early work of \cite{1913ApJ....38..292E} on the \ion{Ca}{ii}~H\,and\,K lines 
was followed up, most famously, by extensive monitoring of the chromospheric activity variability of the same lines by the Mt. Wilson survey \citep{1978ApJ...226..379W, 1995ApJ...438..269B}.
However, the well-known Fraunhofer H\,and\,K lines are not the only spectral lines of \ion{Ca}{ii} that are affected by chromospheric activity. 
In this work, we analyse the \ion{Ca}{ii} infrared triplet (IRT) lines at 8498\,\AA, 8542\,\AA, and 8662\,\AA. 

These lines were first used by \cite{1979ApJS...41..481L} to probe stellar activity because they are formed by the same upper energy level as the H\,and\,K lines \citep{1979ApJS...41..481L, 1989A&AS...80..189F} and hence can be used in a similar manner. 
Their position in the near infrared (photometric $I$ band) enables observations of cool main sequence stars, which have their peak emission in the near infrared, out to larger distances.
In fact, the IRT lines have not only been used to study field stars (e.g. \citealt{1979ApJS...41..481L}, \citealt{2013ApJ...776..127Z}) but also stellar activity in open clusters\footnote{With the future \emph{Gaia}~DR3, many more spectra of the IRT region will become available, enabling activity studies on the Galactic scale, including on numerous open clusters.}.

\cite{2020PASJ...72...80Y} studied the youngest stars in open clusters and star forming regions and found that most of the T-Tauri stars in their sample show magnetically induced activity similar to that seen in older stars. 
\cite{2009MNRAS.399..888M} observed the open clusters IC~2391 and IC~2602, which host F- and G-type zero-age main sequence (ZAMS) stars and pre-main sequence stars of lower mass. \cite{1993ApJS...85..315S} studied the solar-like stars in the ZAMS Pleiades cluster, and \cite{2010MNRAS.407..465J} observed M\,dwarfs in NGC\,2516. 
The oldest open cluster to date with stellar activity measured from the IRT lines is the 220\,Myr old NGC\,6475 (M\,7; \citealt{1997MNRAS.292..252J}). 
Finally, \cite{1993AJ....105..226S} measured stellar activity with the IRT lines for stars in the sparsely populated Ursa Major Moving Group. 
However, its age is not well constrained (\citealt{2015ApJ...813...58J} and references therein). 
In this work we extend the age sequence with the 300\,Myr old open cluster NGC\,3532 and analyse the rotation-activity connection in detail using its rich coeval population.

NGC\,3532 is a very populous open cluster 
that is potentially capable of becoming a benchmark object.
It is located in Carina, at a distance of 484\,pc, and is somewhat embedded in the Galactic disc. 
For a recent detailed overview of NGC\,3532 and related work, we refer the reader to the introduction of \citet[][hereafter F19]{2019A&A...622A.110F}. 
Our prior efforts on NGC\,3532 include a spectroscopic membership study (F19) and, in a companion paper, measurements of the stellar rotation periods of its cool star members (\citealt{2021A&A...652A..60F}, hereafter F21rot). 
Here, we present measurements of the chromospheric activity in this cluster, 
examine their patterns, especially with respect to the rotation periods, and use this knowledge to determine a set of activity-guided photometric rotation periods.

This paper is structured as follows. 
In Sect.~\ref{sec:obs} we present our observations, followed by the measurement of the chromospheric emission in Sect.~\ref{sec:measure}. 
In Sect.~\ref{sec:variability} we compare the chromospheric activity to the photospheric one. 
Afterwards, we combine the chromospheric activity with photometric rotation periods of stars in NGC\,3532 in Sect.~\ref{sec:chromact} to study the colour-rotation-activity connection in detail. 
Here, we construct a colour-activity diagram (CAD) and gain insight into the rotational state of stars without photometric rotation periods. 
In Sect.~\ref{sec:rotact} we construct and analyse the first rotation-activity diagram for NGC\,3532. 
Furthermore, we derive activity-informed photometric rotation periods for members of NGC\,3532.
The rotational implications of these new periods are analysed in detail in the companion paper (F21rot).

%
%

\section{Observations, data reduction, and cluster membership}
\label{sec:obs}
\subsection{Spectroscopy}

We observed NGC\,3532 with the AAOmega multi-object spectrograph on the 3.9\,m Anglo-Australian Telescope (ATT) on 10 and 11 March 2017 primarily from the viewpoint of the membership study presented in F19. 
The spectrograph is fed by the 392-fibre 2dF fibre positioner \citep{2002MNRAS.333..279L}, for which we used three different fibre configurations to observe 1060 stars with the 1700D grating ($R=10\,000$). 
Two of the fibre setups targeted the brighter stars ($I_c=12-15$) with a total exposure time of 30\,min each, each split into three sub-exposures per spectrum. 
The final fibre configuration targeted the fainter stars ($I_c=15-17$) and was exposed for 120\,min.

The data were reduced with the 2dFr pipeline (\citealt{2015ascl.soft05015A}, version 6.28), as described in F19; we use the final data products from that work for the present analysis. 
Before measuring the equivalent widths (EWs), we normalised each spectrum with a fifth-order Chebyshev polynomial. 
In addition to the spectra, we use the measured radial velocity shifts (in pixel) from F19.

\subsection{Photometry}

The time series photometry used in this work was obtained between 19 February 2008 and 1 April 2008 from the Cerro Tololo Inter-American Observatory (CTIO) with the Yale 1\,m telescope operated by the SMARTS consortium. 
We observed eight different fields in NGC\,3532 with three different exposure settings in both the $V$ (120\,s) and $I_c$ (60\,s and 600\,s) filters. 
After standard data reduction, we performed point-spread function photometry with \textsc{DaoPhot~II}. 
The observations, data reduction, photometry, time series analysis, and rotation period determinations are described in the companion paper (F21rot) where these observations are the centrepiece.

For the standardised photometry, we used the optical $BV(RI)_c$ cluster survey of \cite{2011AJ....141..115C} (C11) and combined it with 2MASS \citep{2006AJ....131.1163S} $K_s$ for the primary colour $(V-K_s)_0$ of our analysis. 
The photometry was de-reddened with the mean reddening towards NGC\,3532 of $E_{(B-V)}=0.034$\,mag.

\subsection{Membership and multiplicity}

\begin{figure}
	\includegraphics[width=\columnwidth]{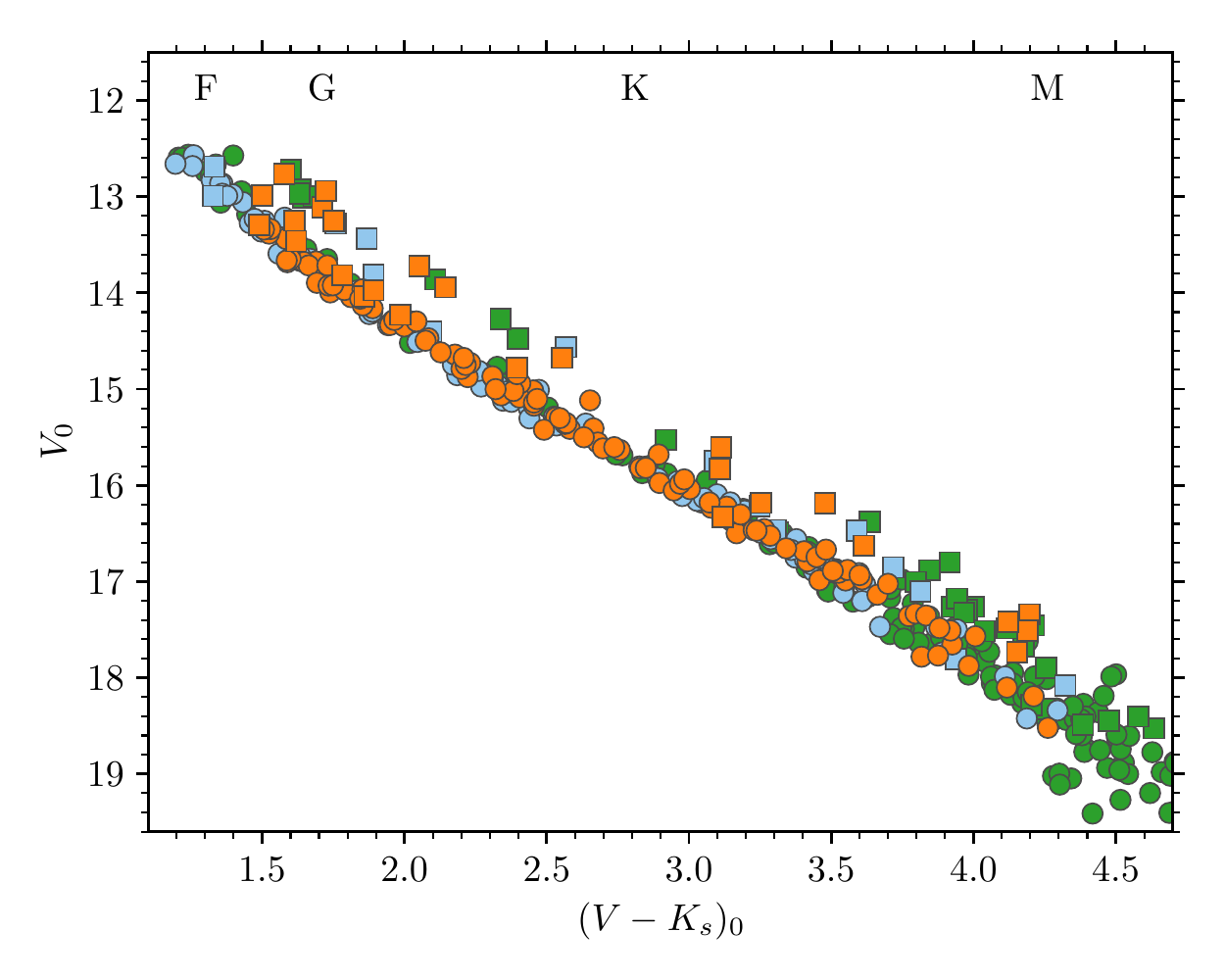}
	\caption{Colour-magnitude diagram of the spectroscopically observed members of NGC\,3532. In orange we show the rotators from \cite{2021A&A...652A..60F} and in light blue from this work. 
We were unable to determine rotation periods for stars marked in green. Single stars are indicated with circles and candidate binaries from photometry and astrometry with squares.}
        \label{fig:CMD}
\end{figure}

The spectroscopic observations presented here were originally obtained to establish a radial velocity membership list for NGC\,3532. 
In F19, we presented a set of 660 exclusive members of NGC\,3532 based on our own observations as well as survey and literature radial velocities. 
Here, we analyse the stellar activity of the exclusive members within our spectroscopic dataset. In Fig.~\ref{fig:CMD}, we show the spectroscopically observed members of NGC\,3532 and highlight different subsets of stars with measured rotation periods (`rotators' hereafter).

Unidentified close binaries could potentially skew our results because binarity can influence a star's magnetic and rotational history (e.g. \citealt{2005ApJ...620..970M, 2007ApJ...665L.155M}).
To assist the analysis of stellar activity and rotation, we searched for potential binaries among the cluster members in the companion paper F21rot. There, we classified cluster members as binaries if their distance to the main sequence locus is $\Delta G > 0.25$\,mag (photometric criterion) or their re-normalised unit weight error (RUWE) from Gaia early data release 3 \citep{2021A&A...649A...2L} satisfies $\mathrm{RUWE}>1.2$ (astrometric criterion, \citealt{2020MNRAS.496.1922B}). In combination, we find 151 potential binaries (22\,\%). These are marked in all relevant figures with squares. 

%
%

\section{Chromospheric emission measurement}
\label{sec:measure}

Previous studies of stellar chromospheric emission in open clusters in the IRT lines (e.g. \citealt{2009MNRAS.399..888M}) used spectra of inactive stars and subtracted them from their target spectra to obtain the chromospheric emission. 
However, this approach is not possible with our dataset because the observations were initially planned as a pure membership survey without the (intentional) inclusion of inactive field stars. 
The number of true main sequence non-members (in radial velocity, proper motion, and parallax) is too small to build a grid of reference spectra. 
Stars that pass one or more of the membership criteria could still potentially be members, meaning young and active, and can certainly not be considered inactive. 
The empirical estimates of the basal flux of inactive stars \citep{2007A&A...466.1089B,2017A&A...605A.113M} were also excluded because of their incompleteness with respect to our sample.

Given the lack of suitable reference spectra, we followed a different approach and measured the EW in a 1\,\AA{}-wide window in each of the three line cores without prior subtraction of a reference spectrum. 
We note that a similar approach is often chosen for the \ion{Ca}{ii}~H\,and\,K lines (e.g. \citealt{1995ApJ...438..269B}) because the tight 1\,\AA{} (triangular) spectral window includes only the line core that contains the signature of the stellar activity -- the chromospheric emission.

\begin{figure}
        \includegraphics[width=\columnwidth]{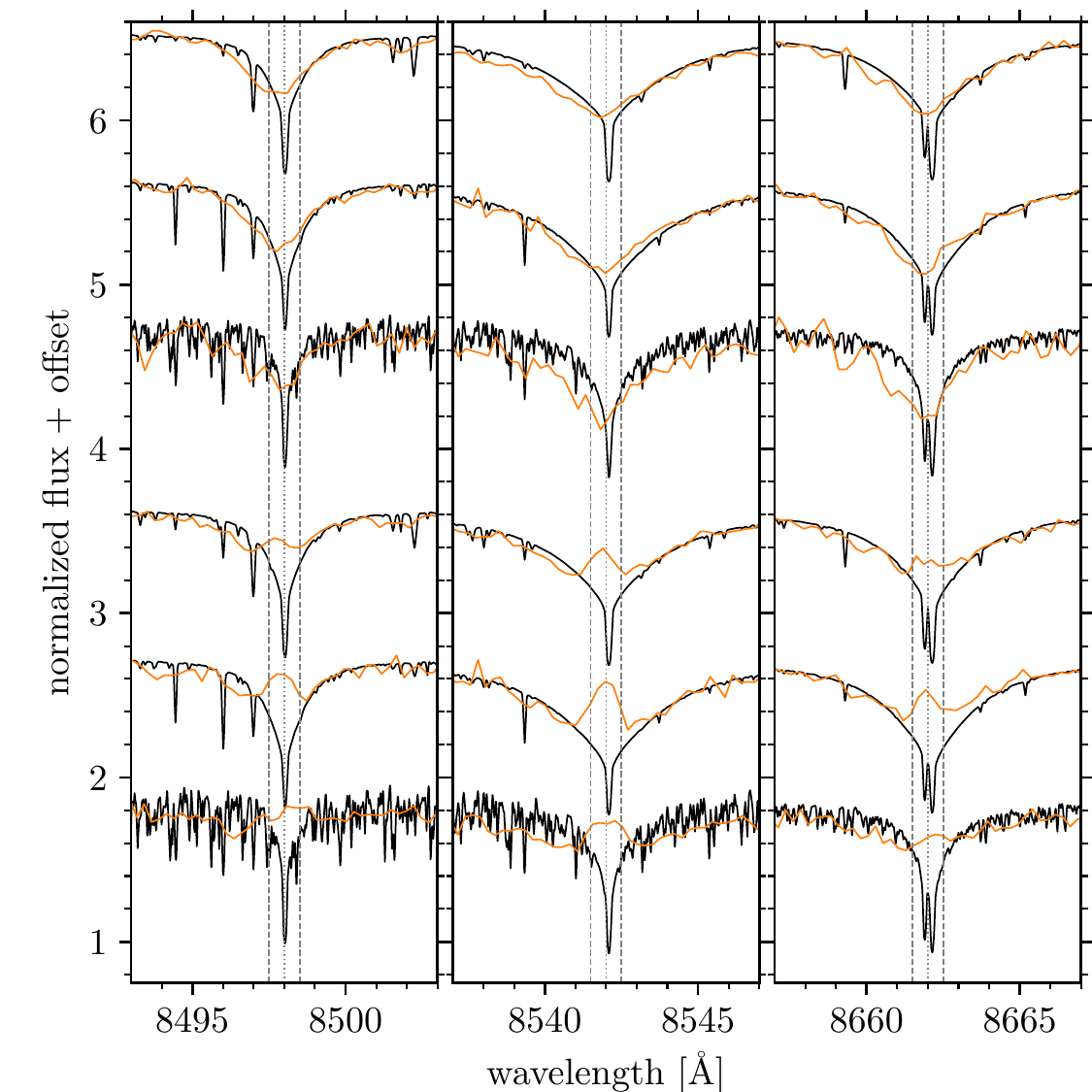}
        \caption{Spectral cutouts for example stars of different chromospheric activity (orange) and their corresponding \textsc{Phoenix} spectra. Each panel is centred on one IRT line and shifted to the rest frame. The line centre is marked with a dotted line and the limits of the 1\,\AA\ wide integration window with dashed lines. The spectra are grouped into low-activity (\emph{top}) and high-activity  stars (\emph{bottom}). The latter feature prominent fill-ins of the line cores. Earlier spectral types are higher. The topmost spectrum is from a G2 star, and the lowest is from an M3 star. All six stars are marked with squares in the colour-activity diagram in Fig.~\ref{fig:IRTactivity}a.}
        \label{fig:spectra}
\end{figure}

The resolution of the AAOmega spectra is too low to simply integrate over a 1\,\AA{} bandpass. 
Therefore, we linearly interpolated between the pixels in a 2\,\AA{}-wide window and oversampled by a factor of ten. 
Thereafter, we placed the 1\,\AA{} bandpass according to the stellar radial velocity shift (Fig.~\ref{fig:spectra}) and integrated over it
to obtain the EW in terms of $\lambda$ and the measured and continuum fluxes

\begin{equation}
EW = \Delta\lambda \sum_{i=0}^n\frac{F_i}{F_{\mathrm{cont.},i}} 
.\end{equation}

We defined the continuum flux $F_\mathrm{cont.}$ with an error-weighted least-squares linear fit to the continuum outside of the IRT lines. 
The fit was evaluated at the position of the integration window, creating a local pseudo-continuum. 
We followed the approach presented by  \cite{2001MNRAS.326..959C} and used their spectral windows.  
\cite{2001MNRAS.326..959C} explain that such a fit improves the continuum determination, in particular in the near infrared with many sky lines and varying signal-to-noise ratio\footnote{Due to the prior normalisation with polynomials, the continuum level is only marginally different from 1 in our case.}.

In contrast to the \ion{Ca}{II}~H\,and\,K lines, the \ion{Ca}{II} IRT lines have a significant photospheric component. 
We are unable to remove this flux using inactive stars observed with the same instrument, as explained earlier. 
Hence, we measured the EW of the IRT lines in the same manner in down-sampled model spectra\footnote{The model spectra in Fig.~\ref{fig:spectra} are displayed in full resolution to give the reader a sense of the spectral features; for all calculations, the spectra have been resampled to the resolution of the observed spectra.} from the \textsc{Phoenix} spectral library \citep{2013A&A...553A...6H}. 
These spectra do not contain a chromospheric component. 
The \textsc{Phoenix} EWs were linearly interpolated over the effective temperature and act as a substitute for our inactive `reference stars'.

To connect this EW scale to our observation, we estimated the effective temperature for each spectrum of the NGC\,3532 members from the relations by \cite{2015ApJ...804...64M} through the $(V-I_c)_0$ colour index and subtracted the photospheric flux (in EW) of the corresponding \textsc{Phoenix} model from the measured EWs. 
The sum of the EW differences of all three lines is defined as the chromospheric emission in the IRT. 
It was converted to an excess flux ($F'_\mathrm{IRT}$) through the empirical flux scale of \cite{1996PASP..108..313H}\footnote{The more accurate empirical flux scale by \cite{2016A&A...595A..11L} is optimised for solar-mass stars and is unfortunately not established for M\,dwarfs.}. 
In addition to the surface flux in the IRT, we calculated the bolometric flux $F_\mathrm{bol}$, assuming a black body, with the Stefan-Boltzmann law. The flux ratio in the IRT follows as (cf. \citealt{2009MNRAS.399..888M}):

\begin{equation}
\log R'_\mathrm{IRT} = \log\left(F'_\mathrm{IRT} / F_\mathrm{bol}\right) = \log F'_\mathrm{IRT} - \left(\log \sigma + 4  \log
T_\mathrm{eff}\right)
.\end{equation}

For each data point, we propagated the resulting uncertainty from the observations. 
In addition, we also include the uncertainty in the estimated effective temperature (80\,K, \citealt{2015ApJ...804...64M}). 
When subtracting the photospheric flux, we also calculated the chromospheric emission of an 80\,K warmer reference and include the difference in the uncertainty. 
The temperature uncertainty was also considered in the calculation of the bolometric luminosity.

In our estimate, we ignored the uncertainty associated with the \cite{1996PASP..108..313H} flux scale for two reasons. 
Firstly, when considering equally massive stars in the colour-activity diagram, the conversion factor from equivalent width to chromospheric flux is constant. 
Hence, no matter how large the uncertainty is, the scatter among the sequence is determined by the measurement of the EW and the intrinsic spread. 
Secondly, we do not find evidence in either the colour-activity diagram or the rotation-activity diagram  that the observed spread is an effect of the flux conversion. 
Hence, we conclude that the uncertainty estimated by \cite{1996PASP..108..313H} is too large\footnote{This is also seen in the good agreement between \cite{1996PASP..108..313H} and \cite{2016A&A...595A..11L}.} and can even be ignored in the case of a coeval group, such as an open cluster\footnote{This is unlikely to be the case when studying field stars because two stars of the same mass can have widely different rotational and activity properties.}.
The IRT excess-EW, the corresponding flux $F'_\mathrm{IRT}$, and the chromospheric emission ratio $\log R'_\mathrm{IRT}$, including their respective uncertainties, are provided in Table~\ref{tab:IRTdata}.

\begin{table*} 
        \caption{Chromospheric activity measurements, including uncertainties and auxiliary data, for members of NGC\,3532. The full table is available online.}
        \label{tab:IRTdata}
        \resizebox{\textwidth}{!}{%
        \begin{tabular}{lllllllllllllll}
                \hline
                \hline
                RAJ2000 & DEJ2000 & C11  & $(V-K_s)_0$ & $T_\mathrm{eff} $& excess EW & $\Delta$EW& $F'_\mathrm{IRT}$ & $\Delta F'_\mathrm{IRT}$ &  $\log R'_\mathrm{IRT}$ & $\Delta \log R'_\mathrm{IRT}$  & $P_\mathrm{rot}$ & $\Delta P_\mathrm{rot}$ & $Ro$ & $\Delta Ro$\\
                (deg) & (deg) & & (mag) & (K) & (\AA) & (\AA) &  ($10^6$\,erg\,s$^{-1}$\,cm$^{-2}$) & ($10^6$\,erg\,s$^{-1}$\,cm$^{-2}$) & & & (d) & (d) & &\\
                \hline
                167.39084 & -58.49086 & 5654 & 4.711 & 3454 & 0.4 & 0.14 & 5.8382 & 0.00031 & -4.068 & 0.071 & \dots & \dots & \dots & \dots\\
                167.37893 & -58.34141 & 6980 & 4.039 & 3748 & 1.2 & 0.04 & 6.4094 & 0.00002 & -3.639 & 0.027 & \dots & \dots & \dots & \dots\\
                167.39199 & -59.16781 & 8911 & 4.000 & 3750 & 1.1 & 0.05 & 6.3632 & 0.00001 & -3.687 & 0.029 & \dots & \dots & \dots & \dots\\
                167.37271 & -58.69355 & 9857 & 3.462 & 4061 & 0.5 & 0.06 & 6.0964 & 0.00002 & -4.092 & 0.037 & \dots & \dots & \dots & \dots\\
                167.35756 & -58.58843 & 11976 & 2.326 & 5028 & 1.2 & 0.01 & 6.7122 & 0.00002 & -3.847 & 0.011 & \dots & \dots & \dots & \dots\\
                167.35757 & -58.89454 & 13437 & 4.171 & 3688 & 0.3 & 0.16 & 5.7948 & 0.00012 & -4.226 & 0.079 & \dots & \dots & \dots & \dots\\
                167.35080 & -58.72758 & 13824 & 1.876 & 5379 & 0.5 & 0.02 & 6.4335 & 0.00001 & -4.243 & 0.014 & 6.40 & 0.96 & 0.1421 & 0.0228\\
                167.33891 & -58.81675 & 16353 & 3.374 & 4094 & 0.7 & 0.04 & 6.2720 & 0.00001 & -3.930 & 0.025 & 10.00 & 0.29 & 0.0943 & 0.0066\\
                167.32970 & -58.63617 & 17130 & 3.982 & 3712 & 0.7 & 0.07 & 6.1779 & 0.00004 & -3.854 & 0.040 & \dots & \dots & \dots & \dots\\
                \dots &&&&&&&&&\\
                \hline
        \end{tabular}}
        \tablefoot{\emph{C11}: ID from \cite{2011AJ....141..115C}; $(V-K_s)_0$ with $V$ from \cite{2011AJ....141..115C} and $K_s$ from 2MASS, de-reddened with $E_{(B-V)}=0.034$\,mag; $T_\mathrm{eff}$ estimated from $(V-I_c)_0$ with \cite{2015ApJ...804...64M}; $P_\mathrm{rot}$ from the companion paper and this work; $Ro=P_\mathrm{rot}/\tau$ with $\tau$ from \cite{2010ApJ...721..675B}.}
\end{table*}

Among the spectroscopically observed stars in the AAO campaign, we find 454 radial velocity members. 
For each spectrum, we are able to measure the chromospheric emission, providing one of the largest sets of chromospheric activity measurements in any open cluster to date\footnote{The largest set we found in the literature contains 516 members of Praesepe for which \cite{2014ApJ...795..161D} measured the chromospheric activity in H$\alpha$.}. 
Out of these 454 stars, 387 stars fall within the field of our photometry, meaning that we have light curves for them. 
We were able to obtain photometric rotation periods for 142 of them, as described exhaustively in the companion paper F21rot. 

These periods enable us to probe the rotation-activity relation in this 300\,Myr old population and to understand the connection between rotation and activity in a mass-dependent manner. 
Furthermore, we are able to increase the number of photometric rotation periods through an activity-informed search in the time series data among the 245 ($= 387 - 142$) remaining stars with light curves but without directly measured prior periods. 
Prior to these analyses (Sect.~\ref{sec:chromact}~\&~\ref{sec:rotact}), we use the light curves of all stars in the spectroscopic sample to probe the connection between photospheric and chromospheric activity.

%
%

\section{Chromospheric against photospheric activity}
\label{sec:variability}

\subsection{Photospheric activity}
Stellar activity manifests itself in various forms and can be observed in all layers of the stellar atmosphere. 
In the photospheric layer, starspots are a widely observed magnetic feature of solar-type stars (e.g. \citealt{2009A&ARv..17..251S}), and are often probed using photometric time series observations.
The simplest measure of such photospheric activity is the variability amplitude of the light curve,
which we can obtain from our own photometric time series.
We use the difference between the fifth and the ninety-fifth percentile of the sigma-clipped light curve \citep{2011AJ....141...20B}. We note that a similar measure is called $R_\mathrm{rng}$ in \cite{2011AJ....141...20B} and measured as a flux difference. In our work we measure it as a magnitude difference and hence call it $V_\mathrm{P95}$ to indicate the difference\footnote{In practice, both versions are nearly identical and can be used in a direct comparison.}.

\begin{figure*}
	\includegraphics[width=\textwidth]{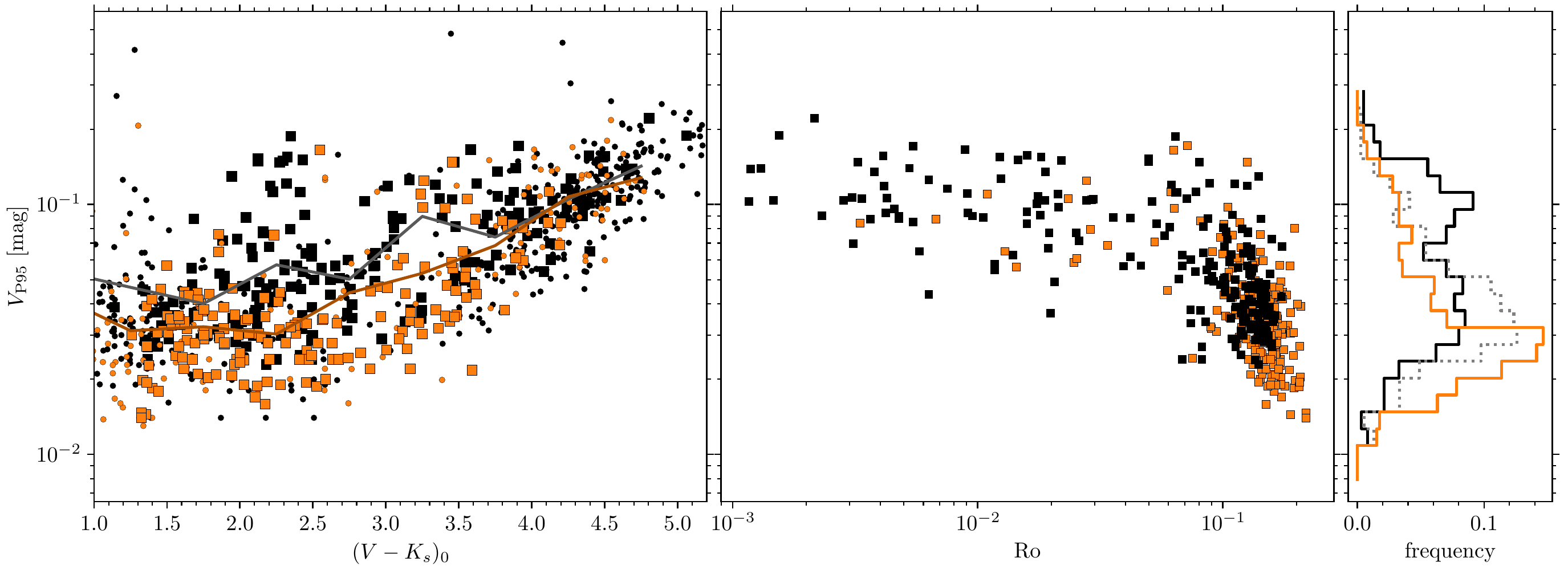}
	\caption{\emph{Left:} $V$ band variability amplitude ($V_\mathrm{P95}$) for stars in NGC\,3532 (orange), compared with those of the zero-age main sequence open cluster NGC\,2516 (black) for both stars with periods (squares) and those without (circles). Stars in NGC\,3532 generally have lower variability, as seen by the mean amplitudes expressed by the solid lines. \emph{Centre:} $V_\mathrm{P90}$ against Rossby number $Ro$ for the rotators in NGC\,3532 and NGC\,2516. The distribution can be seen to resemble the activity-rotation relation previously known from chromospheric and coronal activity, and it shows that the photospheric activity amplitude has a similar structure. \emph{Right:} Histogram of the variability amplitudes for the members of the two open clusters, showing that the high-amplitude stars of NGC\,3532 are far less numerous in comparison with those of NGC\,2516. The dotted grey histogram is restricted to NGC\,2516 stars with $(V-K_s)_0<4$ and confirms that the high-amplitude peak in NGC\,2516 is mostly attributable to the fast-rotating M\,dwarf members, a population absent from our NGC\,3532 data.}
	\label{fig:amplitudes}
\end{figure*}

Because the variability amplitude is a proxy for the photospheric stellar activity, it appears to have similar dependences on stellar rotation and magnetic field as chromospheric and coronal activity measures. 
The left panel in Fig.~\ref{fig:amplitudes} shows the variability amplitude against the intrinsic colour for NGC\,3532. 
(We note that the $(V-K_s)_0$ colour of a Sun-like G2 star is approximately 1.5.)
We removed the bluest stars of our sample because they are overexposed.
In data optimised for such stars, the activity is observed to be low, in keeping with the thinning of convection zones as the Kraft break is approached. In contrast, the higher variability among the lowest-mass stars is intrinsic, and an expression of their faster rotation.

For comparison, we added the variability amplitudes of members in the 150\,Myr-old (ZAMS) open cluster NGC\,2516 (F20) and show the mean amplitude in 0.5\,mag-wide colour bins. We find that the NGC\,2516 stars generally have higher photospheric activity levels as compared with stars in NGC\,3532 except for the very variable M\,dwarfs, where no difference is perceptible.
This shift towards lower variability and activity with increasing age is expected from the stellar spin-down over time and the corresponding decrease of the magnetic field strength.

Indeed, the lower activity of stars in NGC\,3532 is exclusively driven by the slow rotators, as can be clearly seen in the central panel of Fig.~\ref{fig:amplitudes}. 
Here, we show the variability against the Rossby number for the periodic rotators in both open clusters. 
This diagram resembles the well-known rotation-activity correlation from \ion{Ca}{II}~H\,and\,K chromospheric- and coronal activity (e.g. \citealt{1984ApJ...279..763N}, \citealt{2003A&A...397..147P}). 
The stars with the highest variability amplitudes are found in a flat region among the fastest rotators in a mass-independent way -- the saturated regime. 
In this regime, no difference between stars in NGC\,2516 and NGC\,3532 is observed, 
with stars from both clusters occupying the same elevated region across a wide swath in Rossby number.
The situation is different in the regime of larger Rossby numbers ($\mathrm{Ro} \gtrsim 0.06$), where the variability amplitudes are well correlated with the Rossby number,
a behaviour seen in other activity indicators such as coronal X-rays.
This diagram clearly shows that the lower photometric variability 
of stars in NGC\,3532 is a function of the Rossby number and therefore a sign of the slower rotation compared to NGC\,2516.

The lower overall variability can also be seen in the histogram in the rightmost panel of Fig.~\ref{fig:amplitudes}. 
The distribution for NGC\,3532 has the majority of stars between 14\,mmag and 40\,mmag (median amplitude 32\,mmag), with a single-peaked distribution. 
For the younger NGC\,2516, in contrast, we find a double-peaked distribution between 20\,mag and 120\,mmag (median 67\,mmag). 
The double-peaked structure can be traced to the prevalence of many fast-rotating M\,dwarfs in NGC\,2516. 
If we restrict the colour range to $(V-K_s)_0 < 4$ (thereby removing the M\,dwarfs), the peak at the higher amplitudes disappears. 
However, even in this restricted sample the median amplitude for NGC\,2516 is higher than that for NGC\,3532.

\begin{figure}
        \includegraphics[width=\columnwidth]{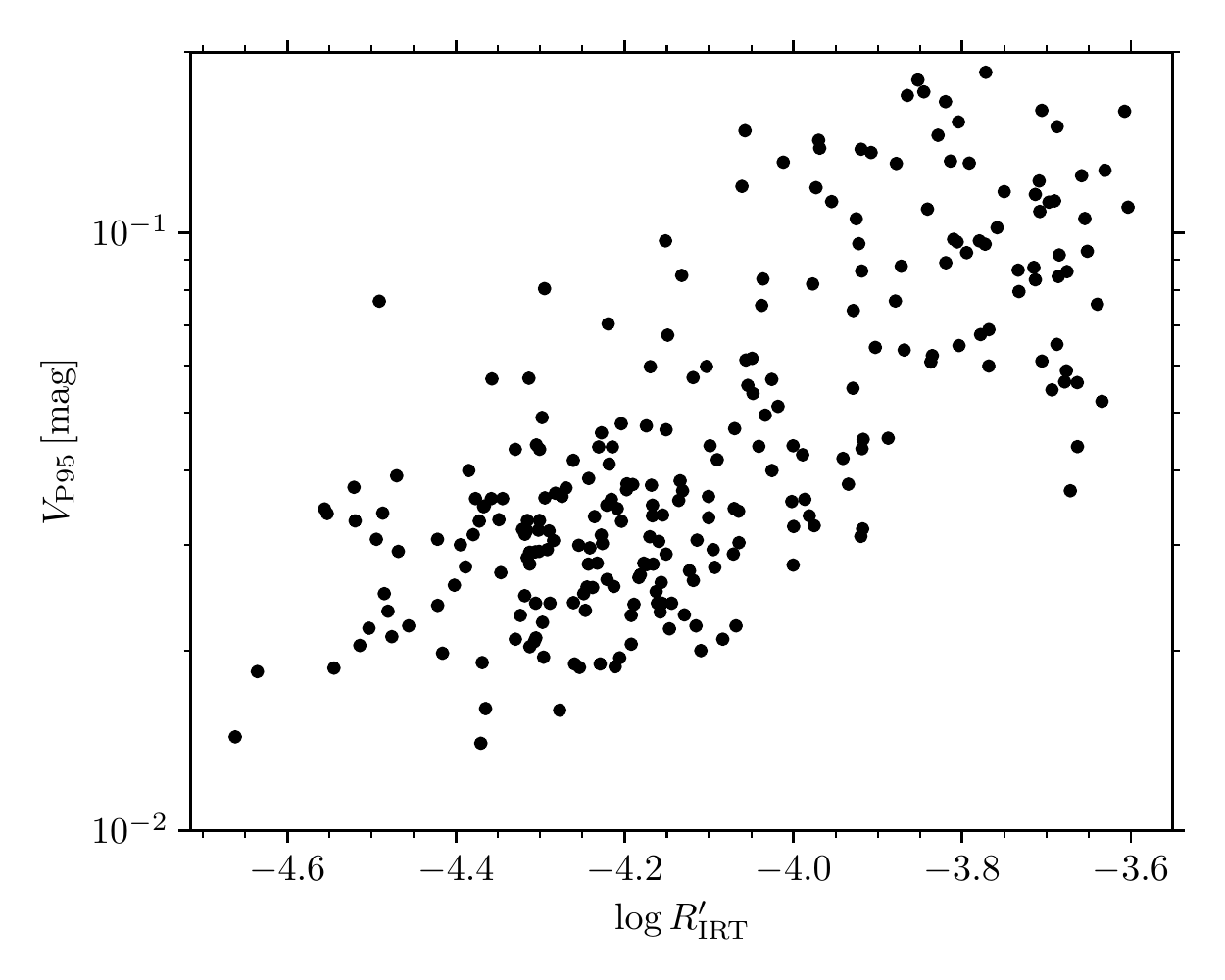}
        \caption{Photometric variability amplitude in the $V$ band $V_\mathrm{P95}$ against chromospheric emission ratio $\log R'_\mathrm{IRT}$. The two activity indicators are correlated and represent two facets of the underlying magnetic activity.}
        \label{fig:amps}
\end{figure}

\subsection{Connection between photospheric and chromospheric activity}

Because the photospheric activity shows the same structure as is observed in chromospheric activity, we expect the two to be correlated. 
In Fig.~\ref{fig:amps}, we therefore show the photometric variability amplitude in the $V$ band $V_\mathrm{P90,V}$ against the chromospheric emission ratio $\log R'_\mathrm{IRT}$. 
A clear trend (correlation coefficient $r=0.665$) is visible in this comparison, with higher photometric variability visible at higher chromospheric flux. 
The large scatter in the trend can potentially be traced to short-term activity observed in the light curves\footnote{A longer baseline timescale could reduce this scatter.}. 
Additionally, the inclination angle is expected to influence the variability amplitude. 
We note that the spectroscopic and photometric data are not even nearly contemporaneous, but were obtained nine years apart. 
Despite this significant time difference, the chromospheric and photospheric activity are
well correlated, not unexpected because of the underlying common origin -- the stellar magnetic field and the stellar rotation, both of which evolve on much longer timescales.

In related work, \cite{2020ApJS..247....9Z} have compared variability amplitudes of stars in the \emph{Kepler} field to their Ca~H\,and\,K chromospheric emission. 
They found similar results but were able to compare stars of different spectral types because their sample includes stars of various activity levels caused by the inhomogeneous stellar ages in the \emph{Kepler} field. 
This kind of comparison is not possible in our coeval dataset because all F and G\,dwarfs have very similar chromospheric activity. \cite{2018ApJ...855...75R} found a similar correlation between chromospheric activity and photometric activity for young and active stars. Other work on the temporal correlation between photospheric and chromospheric activity found anti-correlations \citep{2018AJ....156..203M} and a comparison between two chromospheric activity indicators found a correlation only for very active stars \citep{2006PASP..118..617R}.

%
%

\section{Chromospheric activity and stellar rotation in the colour-activity diagram}
\label{sec:chromact}

\begin{figure*}
	\includegraphics[width=\textwidth]{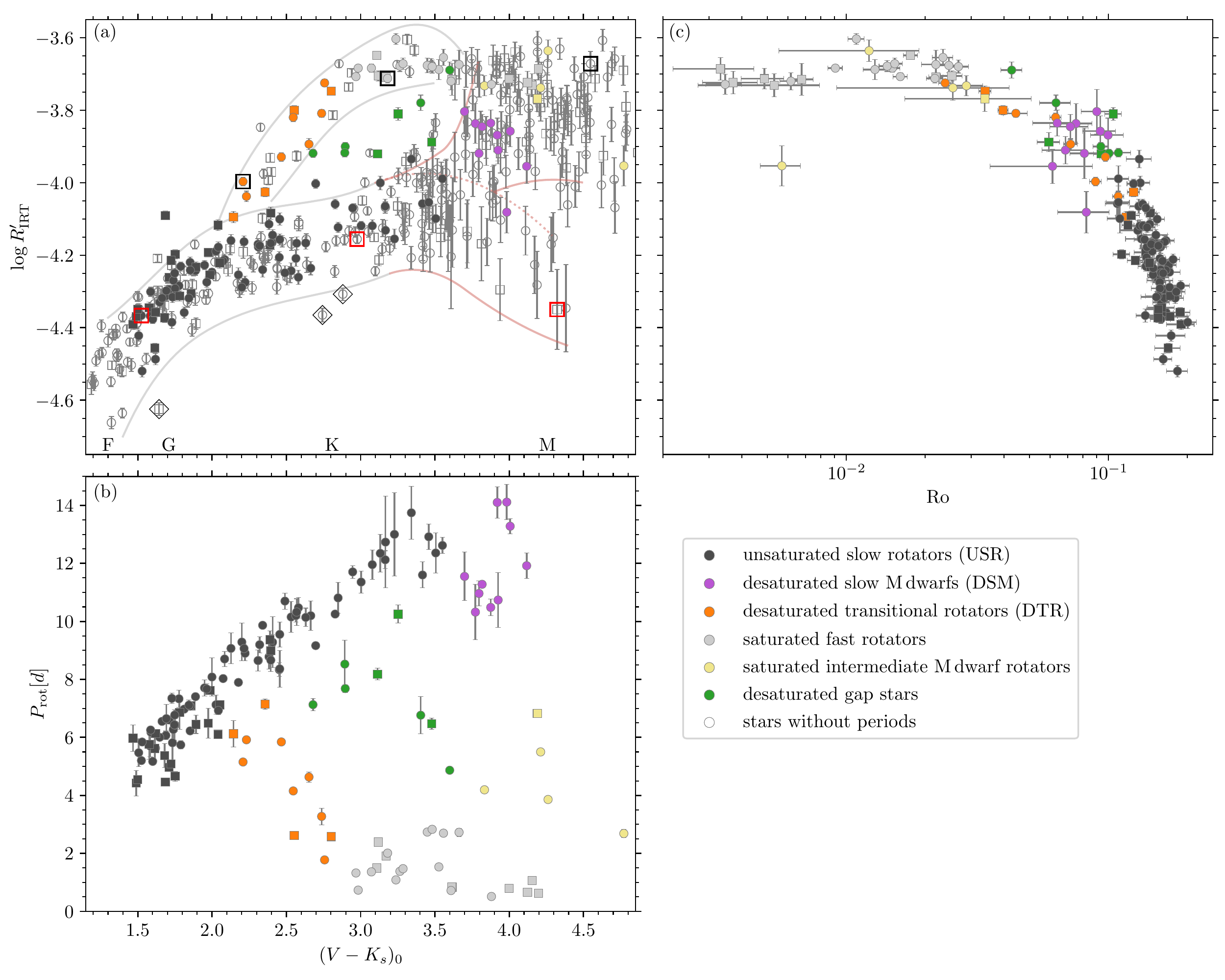}
	\caption{Interconnection between chromospheric activity, rotation, and mass. The colour-activity diagram (\emph{a}) connects the colour-period diagram (\emph{b}) with the rotation-activity diagram (\emph{c}).
	\emph{(a)}: Chromospheric emission ratio $\log R'_\mathrm{IRT}$ against intrinsic $(V-K_s)_0$ colour for members of NGC\,3532. Member stars with measured rotation periods are shown with filled symbols and those without photometric rotation periods with open symbols. Likely binaries are indicated with squares, whereas single stars are marked with circles. The majority of the data points delineate very clear sequences, indicating the cohesion of chromospheric activity evolution within the coeval stars of an open cluster. The colour coding of the different groups is indicated in the legend on the lower right and is the same in all three panels. The thin lines indicate the sequences described in the text. The stars emboxed in red and black squares are the low-activity and high-activity stars, respectively, shown in Fig.~\ref{fig:spectra} (top to bottom in Fig.~\ref{fig:spectra} corresponds to left to right in this figure). The bluest of these stars has roughly the Solar colour. Stars marked by diamonds are outliers (discussed in the text).
	\emph{(b)}: Colour-period diagram of NGC\,3532, with groups of stars distinguished by colour. This diagram is nearly a mirror image of the colour-activity diagram, with higher-activity stars being clearly mapped one-to-one onto the faster rotators. 
	\emph{(c)}: Corresponding rotation-activity diagram, showing the  chromospheric emission ratio against the Rossby number $Ro$. The rotation-activity diagram is remarkably clean despite the diversity in chromospheric activity.}
	\label{fig:IRTactivity}
\end{figure*}

Because chromospheric activity is closely linked to the stellar rotation rate, we study both properties jointly in Fig.~\ref{fig:IRTactivity} to emphasise this connection. 
Our large number of measurements and our provision of rotation periods (instead of $v \sin i$ measurements) will be seen to allow more detailed connections between IRT activity and the relevant quantities than was possible in the most notable prior cluster studies to date, those of the Pleiades \citep{1993ApJS...85..315S} and NGC\,6475 \citep{1997MNRAS.292..252J}.

In panel a, we display the measured activity $\log R'_\mathrm{IRT}$ against the intrinsic $(V-K_s)_0$ colour to investigate its mass dependence\footnote{Readers might consider comparing this with the CAD in Fig.~5 of \cite{1997MNRAS.292..252J}, which shows a similar, but noisier, correlation.}. 
The second diagram in panel b features the colour-period diagram (CPD) from the companion rotational study. The final panel c shows the rotation-activity diagram for NGC\,3532\footnote{A suitable prior comparison for this panel is Fig.~19 of \cite{1993ApJS...85..315S}.}. 
Stars with common features in panel b have been grouped together and colour-coded so that their locations in all three panels can be recognised. This enables the detailed analysis of the rotation-activity connection even in the colour-activity diagram (CAD), which otherwise does not contain the rotational information.

The CAD in Fig.~\ref{fig:IRTactivity}a is analogous to a colour-magnitude diagram (CMD) or a CPD, and it shows the chromospheric emission ratio against the intrinsic colour. 
Our CAD also contains a large number of NGC\,3532 cluster members for which we were unable to derive rotation periods (unfilled symbols).
As in a CMD or CPD the (main sequence) open cluster members form narrow bands in the CAD,
but notably, what was previously believed to constitute a `dispersion' in activity values here appears to resolve into distinct sequences and gaps that mirror the behaviour in the CPD. 
The most prominent of these sequences runs diagonally in the CAD from low stellar activity levels in the warmest (F) stars to high levels among the M\,dwarfs. 
From this main feature a branch of higher-activity stars splits off at $(V-K_s)_0\approx 2.1$. 
At the red end of the CAD, among the lowest-mass stars, we observe a large spread in the activity levels. A similar spread is observed in the stellar rotation periods.
We discuss each group of stars separately below.

\subsection{Unsaturated slow rotators} 

With the help of the colour-coding, the main feature in the CAD can easily be associated with the slow rotator sequence in the CPD (Fig.~\ref{fig:IRTactivity}b). 
There is a very well-defined sequence of FGK\,dwarfs with $\log R'_\mathrm{IRT}\lesssim -4.1$ in the CAD that mirrors the clean slow rotator sequence of such stars in the CPD (dark grey).
We therefore call these stars the `unsaturated slow rotators' (USRs).
It is strengthened considerably by additional stars for which we have been unable to derive rotation periods (open symbols).
In agreement with prior work on the rotation-activity connection, the slow rotators clearly correspond to the lowest-activity stars in each mass bin in the open cluster. 
In keeping with this idea, we identify the M\,dwarfs with $\log R'_\mathrm{IRT} \lesssim -4.2$ as the continuation of the sequence of unsaturated stars. 
These will be found later to lie on the extended slow rotator sequence in the CPD.
At this stage, the classification of the M\,dwarfs marked in purple is ambiguous. We discuss them separately below, and argue below that they should not be considered as USRs because of their increased activity (although short of saturation level) and shorter rotation periods. 
We refer to them here as `desaturated slow M\,dwarfs' (DSMs).
The significantly lower activity of the warmest (F and G) stars relative to the K\,dwarfs, is also unsurprising because they have far shallower surface convection zones.

Three stars (C11: 70134, 72069, 121200, marked with black diamonds in Fig.~\ref{fig:IRTactivity}a.) fall below the well-defined unsaturated sequence. 
Given the correlation between rotation and activity, those stars might be expected to rotate much slower than other equally massive stars in the cluster, implying that they are likely field stars of an older age. 
However, all three stars are bona fide cluster members in radial velocity, proper motion, parallax, and photometry. 
Unfortunately, their light curves and periodograms are too noisy to determine their photometric rotation periods.

 We find only one star with an elevated activity level 
 ($\log R'_\mathrm{IRT}=-4.1$ at $(V-K_s)_0=1.7$)
that places it above this well-defined sequence for the F and G\,dwarfs in the CAD. 
Our measured rotation period, however, places it in agreement with the other slow rotator stars in the CPD (rather than much faster, for instance). 
This star is a photometric binary, and it is possible that binary-induced activity is responsible for its elevated position in the CAD. 
Alternatively, we could have captured it by pure chance during a phase of increased stellar activity. 

\subsection{Desaturated transitional rotators and activity gap}

We have identified an unexpectedly tight sequence of G-K stars with activity 
above the USR level, but below the saturated level.
As can be seen from Fig.~\ref{fig:IRTactivity}a, near $(V-K_s)_0 \gtrsim 2.0$, $\log R'_\mathrm{IRT}\gtrsim -4.1$ a branch of higher-activity stars emerges from the unsaturated sequence. 
Cross-comparison of this higher-activity branch with the CPD reveals these stars to be associated with the stars in transition from fast to slow rotation (marked in orange). 
In fact, the association is one-to-one for stars with measured rotation periods.
This branch of higher-activity stars reaches the saturation level\footnote{The saturation observed in the chromospheric emission is the same as observed in the coronal X-ray emission.} 
at $(V-K_s)_0\approx 3.0$, 
at which point all of its stars are classified as fast rotators (light grey) in the CPD.
We propose the name `desaturated transitional rotators' (DTRs) to capture their transitional rotation-activity behaviour.
There are four binaries (squares) in this group, but the majority (seven) of these stars show no evidence of binarity (circles).

In the gap between the two sequences in the CAD discussed above, we find a small number of stars (marked in green) that correspond exactly to the stars in the rotational gap of NGC\,3532. 
With this knowledge, we can clearly define the rotational gap in the CPD separating the sequences.
In fact, the correspondence with the CPD, where these stars are in transition from fast to slow rotation, tells us that these stars in the CAD are in transition from the high-activity branch to the low-activity one.
In keeping with prior terminology for the rotational gap, and also to reflect the intermediate activity level of these stars, we refer to them as `desaturated gap stars'.

We note that the DTRs have a higher binarity fraction compared to USRs of the same colour. Yet, as the rotational analysis in the companion paper shows, these stars do not spin faster due to binary tidal interactions (as for example in Praesepe; \citealt{2019ApJ...879..100D}) but are the remainder of the initial fast rotators that emerge from the star formation process. In the companion paper (F21rot), we interpret the higher binarity fraction among the fast rotators with the shorter disc lifetime in binary systems. Even so, only  a minority of the DTRs show any sign of binarity, and the higher activity with faster rotation appears not to be directly caused by binarity.

The absence of additional (elevated) stars bluewards of the transitional DTR sequence in the CAD (orange) is also notable, and very important in understanding angular momentum evolution because it clearly shows that even stars without determined photometric rotation periods fall onto the low activity and slow rotator branch in this colour range.
Conversely, this correspondence assures us that all fast rotators down to a certain mass limit have already transitioned to the slow rotation phase. 
The lack of fast-rotating stars in the CPD for solar-mass stars is therefore astrophysical, and not an observational bias caused by small variability amplitudes.

Given the accurate correspondence between the CAD and the CPD, the branching-off point 
of the DTRs
could potentially be used to age-rank young open clusters with high relative sensitivity. 
As shown in the companion rotational study, the transition from fast- to slow rotation happens on short timescales (tens of megayears), and hence is very age-sensitive. 
Using a CAD would correspondingly allow potentially accurate age-ranking to be obtained from a single multi-object spectroscopic observation\footnote{Unlike $v\sin i$, which can also be obtained from a single observation, our method is also very sensitive to slow rotators.}, rather than a time series, as needed for photometric rotation periods.

\subsection{Diversity in activity of low-mass stars}

As one approaches the lower-mass stars towards late K and early M\,dwarfs ($(V-K_s)_0 \gtrsim 3.2$), the activity of the stars is more diverse, and the structure of the CAD becomes correspondingly more complicated.
The diversity is not related to multiplicity as seen from Fig.~\ref{fig:IRTactivity}(a) where the fraction of likely binaries in this regime is seen to be comparable to the USRs.
We discuss this mass regime (indicated by the red outlines in Fig.~\ref{fig:IRTactivity}a) sequentially from high to low activity.

At the top of the CAD, in the saturated regime 
($\log R'_\mathrm{IRT}> -3.8$), we find two corresponding groups in the CPD: the fast rotators (light grey) and a group of stars rotating slightly more slowly  (yellow). 
Despite the latter stars rotating more slowly than the former, their chromospheric activity is still saturated. 
The large number of M\,dwarfs with saturated activity and as yet undetected periods must therefore all belong to one of these two groups. 
We infer that we likely did not have the photon sensitivity to directly detect periods for them.

In a band with $-3.8 > \log R'_\mathrm{IRT} > -4.0$, we find a group of lower-activity stars that are also significantly slower rotators than the saturated activity ones.
These are grouped tightly together in both the CAD and the CPD, where they are marked in purple. 
We call these the `desaturated slow M\,dwarfs' (DSMs) to reflect both their reduced activity levels as compared to their saturated counterparts, and their location in the CPD, near the slow rotator sequence, if not quite on it.
Because of these joint intermediate behaviours, we suggest that these stars are distinct from the low-activity M\,dwarfs (discussed immediately below), and that they should not be considered as a continuation of the USR sequence in the CAD.
That continuation instead is provided by the sizeable group of low-activity M\,dwarfs below them in the CAD.

Finally, we find a group of M\,dwarfs with $\log R'_\mathrm{IRT}< -4.0$. 
They form a sequence of declining activity with declining mass (below the red dotted line).
This sequence connects smoothly to the USRs (dark grey), and we identify these stars as the continuation of the USR sequence into the M\,dwarf regime. 
At this stage of the discussion, we have no rotation periods for any of them, and thus no counterparts for them in the CPD. 
We have likely ignored their rotation periods with our demand for a clearly recognisable rotation period. 
(We reappraise this in the following section and derive their photometric rotation periods, thereby justifying the classification here.)

A sparsely populated gap appears between the two sequences traced among the M\,dwarfs.
Our data are not numerous enough to provide sufficient insight into 
the reality of such a gap
between the low-activity M\,dwarfs and the more active stars. 
However, we note that a gap is present between the fast rotators and the slow rotators among the M\,dwarfs in the ZAMS open clusters Pleiades and NGC\,2516 \citep{2016AJ....152..113R, 2020A&A...641A..51F}, a hint of which is found in the CAD.
\ion{Ca}{II}~H\,and\,K and H$\alpha$ observations of the Pleiades show a similar feature \citep{2018MNRAS.476..908F}. 
It should also be noted that \cite{2018MNRAS.476..908F} have questioned the Pleiades membership of the least active stars despite the very slowly rotating stars already known in the Pleiades \citep{2016AJ....152..113R}.

A gap between slow and fast rotators also exists in field M\,dwarfs. 
The chromospheric emission measurements $(L_{\mathrm{H}\alpha}/L_\mathrm{bol})$ presented by \cite{2017ApJ...834...85N} show two distinct groups of high and low activity. 
They correspond to rapidly and slowly rotating stars that are separated by a gap \citep{2016ApJ...821...93N}. 
With the gap also present in the field star population, it should be well defined and very sparsely populated in open clusters. 
However, more observations of mid-M stars are needed to constrain its extent.

%
%

\section{Rotation-activity relationship and additional rotation periods}
\label{sec:rotact}

Having discussed the mass dependence of the observed stellar activity in the CAD, 
we move on to investigating the dependence of the IRT activity on rotation. 
The wide difference in activity between stars with the same rotation periods (for instance the USRs and the DTRs with similar rotation periods of $4-6$\,d) shows that this dependence is not direct, and that a star's mass (or suitable proxy) has to be considered as well.
Such comparisons are therefore typically performed using a mass-normalised version of the rotation period called the Rossby number, to compensate for the variation in activity with stellar mass.
The first such diagram was devised and presented by \cite{1984ApJ...279..763N}, and has since been constructed not only for the chromospheric emission of solar-type stars, but also for their coronal emissions (e.g. \citealt{2003A&A...397..147P}) and also for the activity of fully convective stars (e.g. \citealt{2011ApJ...743...48W}).

However, a drawback of the Rossby number is that it depends on the convective turnover timescale, which is not directly accessible from observations, and one has to rely on stellar models or empirically calibrated relations. 
For this work, we use the relation between $T_\mathrm{eff}$ and the convective turnover timescale from \cite{2010ApJ...721..675B} to derive it for our stars. 
The effective temperature has already been estimated in Sect.~\ref{sec:measure} through the relationship given by \cite{2015ApJ...804...64M}. 
The uncertainty in $\textrm{Ro}$ was estimated from the period uncertainty and the corresponding calculation for $\tau_c$ using offsets of $\Delta T_\mathrm{eff}=80$\,K. 
As with the uncertainties for $\log R'_\mathrm{IRT}$, this approach only includes the measurement and not model-dependent uncertainties.
The latter typically differ by only a scaling constant, implying that the underlying power-law dependences are unchanged.

\subsection{The rotation-activity relationship}

Upon transformation to $\textrm{Ro}$, 
we find that all stars with measured rotation periods follow a remarkably tight rotation-activity relationship (Fig.~\ref{fig:IRTactivity}c). 
The separate groups and even sequences of stars in the CPD and CAD collapse onto a single relationship, obscuring their diverse behaviours in those diagrams.
Ours echoes similar relationships seen in chromospheric \citep{1984ApJ...279..763N} and coronal activity \citep{2003A&A...397..147P}.
Two regimes known from prior work are instantly recognisable.
In the saturated regime of low Rossby numbers, the chromospheric emission 
is independent of the rotation period and has a value of $\log R'_\mathrm{IRT} \sim -3.7$. 
After a curved transition region (in this log-log representation) in the vicinity of $Ro \approx 0.1$, the chromospheric emission declines steeply with $Ro$ to define the unsaturated regime, where activity depends strongly on rotation.

Our prior division of stars in the CPD into distinct groups provides further insight into the different activity regimes. 
While it is unsurprising that the warmest of the slow rotators are the most evolved from an activity standpoint, and that the fastest M\,dwarfs are all saturated and thus least evolved, we see from the colour-coded divisions made earlier that there is indeed an order in which the activity evolves.
The first stars to evolve after the unsaturated slow rotators (dark grey) are the desaturated slow M\,dwarfs (purple) and the activity gap stars (green), 
finally followed by the saturated intermediate rotators (yellow), and then the saturated fast rotators (light grey). 
Perhaps unsurprisingly, the desaturated transitional rotators (orange) 
 which span a large range of Rossby number, stretch in the rotation-activity diagram all the way from the unsaturated to the saturated stars.

\begin{figure}
        \includegraphics[width=\columnwidth]{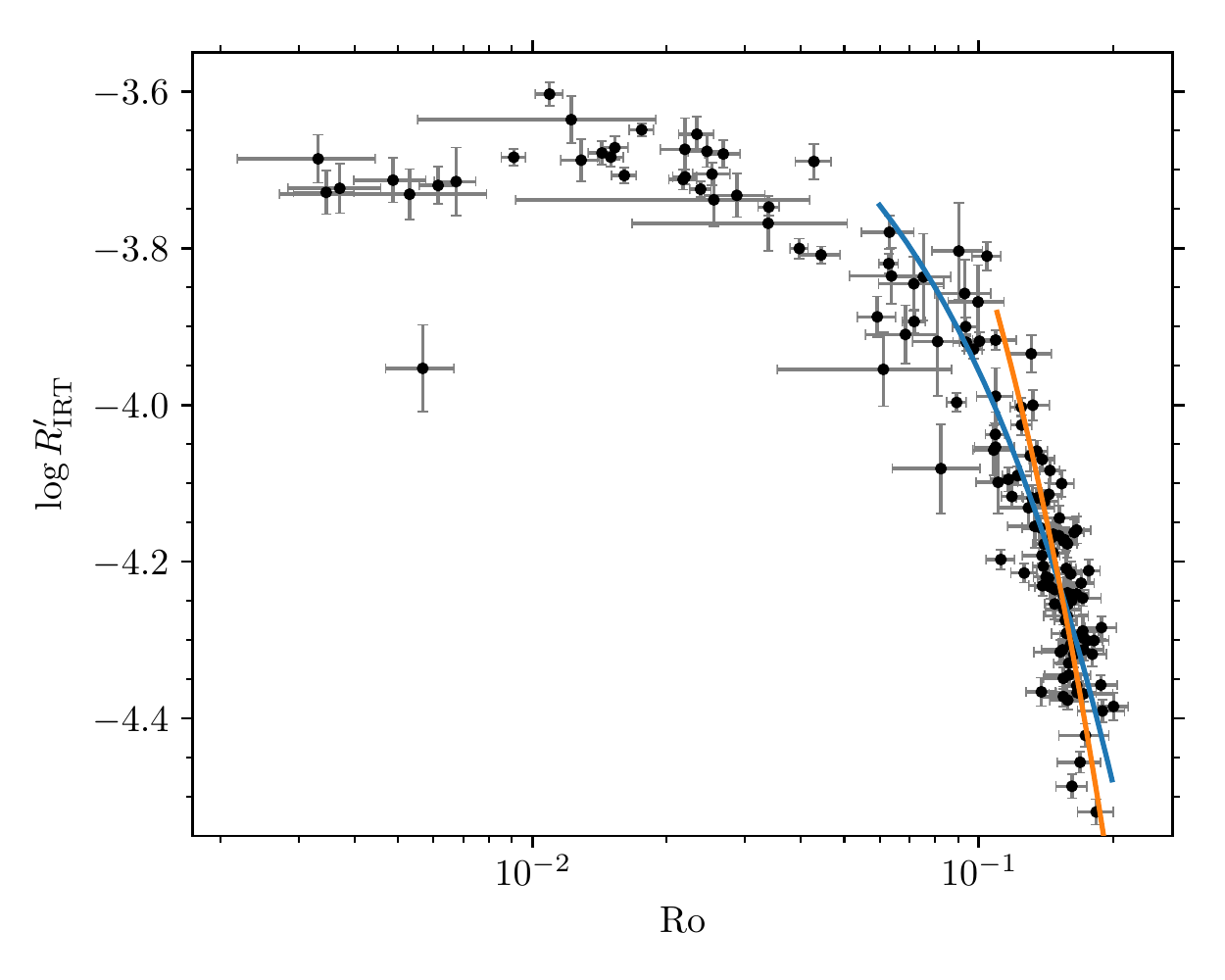}
        \caption{Rotation-activity diagram with two possible log-linear fits to two regions of interest: all unsaturated stars (blue, Ro $> 0.06$) and only the slow rotators (orange, Ro $> 0.11$).}
        \label{fig:RoAct}
\end{figure}

The correlated, unsaturated, part of the rotation activity relation is sometimes described as a power-law (e.g. \citealt{2003A&A...397..147P, 2011ApJ...743...48W}), and at other times with a log-linear function \citep{2008ApJ...687.1264M}. 
We find the log-linear function to follow our data better than a power-law.
This suits our goal of effectively locating the stars without measured rotation periods in the activity-rotation space. 
We fit the correlation in the unsaturated regime with a linear function of the form $\log R'_\mathrm{IRT}= a + b\,Ro$ using Orthogonal Distance Regression\footnote{Unlike ordinary least-squares, ODR includes the uncertainties in both variables.} (ODR; \citealt{ODR}) from the \textsc{SciPy} package.
We perform the fits over two ranges, once for the slow rotators alone (defined as having $Ro > 0.11$) and once for all unsaturated stars (those for which $Ro > 0.06$).
We show both results in the rotation-activity diagram in Fig.~\ref{fig:RoAct}.

For the slow rotators alone, we find 
\begin{equation}
        \log R'_\mathrm{IRT}= (-2.970\pm0.134) + (-8.293\pm0.906) \, \textit{Ro}
        \label{eq:ro11}
,\end{equation}
and including all unsaturated stars the relation is
\begin{equation}
        \log R'_\mathrm{IRT}= (-3.429\pm0.029) + (-5.276\pm0.222) \, \textit{Ro}.
        \label{eq:ro06}
\end{equation}

The root of the differing slopes for both fits might be found among the least active stars. 
Due to small numbers, we find the scatter among these stars to be smaller compared to other regions of the rotation-activity diagram. 
Yet the fit to the slow rotators ($Ro>0.11$) matches these data points because it has fewer constraints among the more active stars. 

\subsection{Lower activity among M\,dwarfs?}

All stars, with only one exception, are seen to follow the activity-rotation relation.
That star (ID 707560, C11: 210716) falls 0.3\,dex (well below) the saturated branch.
It has a short rotation period ($P_\mathrm{rot} = 2.76$\,d) that is associated with a clear peak in the periodograms, and shows no alias periods with significant power. 
The light curve and the periodograms can be found in  Fig.~\ref{fig:LCfullyconvective}. 
Given such properties, we would normally expect it to fall into the saturated regime. 

However, the star is peculiar in more than its activity; 
it is also the lowest-mass rotator in our sample.
Unfortunately, we were not able to derive rotation periods for other stars of similar mass and activity. 
Indeed, a large spread in activity can be found among the least-massive stars in the CAD. Given the connection between rotation and activity, we would also expect a correspondingly large spread in periods. 
However, 707560 is already among the least active stars in the fully convective regime ($(V-K_s)_0 \gtrsim 4.6$), with a rotation period significantly faster than for other early M\,dwarfs of similar activity.

In conclusion, we cannot specify whether this star has an unusually low activity or whether such low-mass stars simply do not follow the rotation activity relation of the higher-mass stars. 
\cite{2010MNRAS.407..465J} have found a decline of the activity, measured in the IRT, with colour for mid-M\,dwarfs in NGC\,2516. 
In their work mid-M\,dwarfs populate an unstructured rotation-activity diagram. 
Such behaviour was not observed by \cite{2014ApJ...795..161D} who use H$\alpha$ as their tracer of chromospheric activity. Similarly, again using H$\alpha$, \cite{2017ApJ...834...85N} find the rotation-activity relation to be valid for mid- and later-M\,dwarfs. 
These stars, however, are field stars and likely not young. 
Unfortunately, our rotation periods do not extend to lower masses and thus do not allow us to study the activity-rotation relation for mid-M\,dwarfs or to probe whether the unstructured rotation-activity diagram from \cite{2010MNRAS.407..465J} is intrinsic to the chromospheric emission in the IRT.

\subsection{Activity-informed period search}
As already pointed out by \cite{1984ApJ...279..763N}, the relation between chromospheric activity indicators and Rossby number in principle allows one to estimate the stellar rotation period (e.g. \citealt{2004ApJS..152..261W}). 
We are in the advantageous position of possessing light curves of most of the stars in our activity sample. 
This allows us not only to estimate the rotation period from the activity, but also subsequently to determine the actual photometric rotation period from the photometry.

The estimated rotation period serves as a prior to our period search,
and we calculated it for every star that falls either onto the slow rotator branch of the CAD ($\log R'_\mathrm{IRT}<-4.0$) or in the transition and gap region ($\log R'_\mathrm{IRT}<-3.8$ and $(V-K_s)_0<3.6$). 
We excluded the saturated stars (mostly M\,dwarfs) for which the activity level obviously cannot provide a useful prediction for the rotation period. 
We calculated the estimated rotation period $P_\mathrm{est} = \textrm{Ro} \, \tau_c$ from Eqs.~\ref{eq:ro11}~and~\ref{eq:ro06}, respectively, for the two samples. 
The spread in Rossby number among the stars on the slow rotator sequence is ${\sim}10\,\%$. 
The same relative uncertainty is assigned to $P_\mathrm{est}$ as $\Delta P_\mathrm{est}$. 

As described exhaustively in the companion paper,
we calculated periodograms for each light curve with five different methods. 
However, for this particular phase of the work we use only four, discarding the string length method on account of its periodogram becoming noisy for low S/N light curves. 
Additionally, this method often does not agree with the other four methods, unnecessarily inflating our uncertainties. 
In the periodograms, we search for the highest peak within $2.5 \Delta P_\mathrm{est}$ of $P_\mathrm{est}$. 
A manual selection process verifies the existence of a periodic signal in the light curve with this period. A visualisation of this process is given with the example light curves in Fig.~\ref{fig:exampleLC}. We show the estimated period and the search range in the different periodograms. In keeping with the period determination algorithm in the companion paper, we calculate the final period as the mean of all four possible values, and its uncertainty as the spread among the four initial values.
During this process, we restricted the search range for ${\sim}15$ light curves to assign the activity-guided period, and not a (chance) higher secondary or alias peak in the periodogram. 
We confirm all of our previously determined rotation periods, showing that we have included no alias periods among the slow rotators.
Additionally, we find 116 
new rotation periods among stars previously classified as non-rotators. 
By definition all of these rotation periods follow the rotation-activity relation. 
We provide the newly found periods in Table~\ref{tab:IRTdata} along with the chromospheric activity data.

The success of this method can be attributed to multiple reasons. 
Firstly, the stellar activity measurement narrows the possible period range and enables us to choose the correct peak in noisy periodograms. 
It gives us confidence in assigning a period to a noisy light curve and enables us to find the true period in light curves with evolving spots. 
A second major reason is the breaking of aliases. 
Since we know that all of our stars are slow rotators (or in the rotational gap), the rotation period cannot be near 1\,d for instance, and hence we break the alias with the observational cadence. 
Finally, we are also able to break the half and double period alias because a factor of two is a very large offset in the rotation-activity diagram. 
Jointly, these reasons increase the yield of rotation periods for our
ground-based observations from 37\,\% to 66\,\% (255/387 stars\footnote{Excluding three potential non-members among the M\,dwarfs (see Sect.~\ref{sec:Mdmem}).} with AAO spectra) in the correlated regime.

However, we are unable to assign a rotation period to every light curve in this set of slow rotators.
Among the 184 considered stars, we were not able to assign a rotation period to nine.
In some cases the light curve is just too noisy, resulting in periodograms without distinctive peaks. 
Occasionally, although peaks are found, the different methods provide inconsistent results. 
In both cases, we do not assign a rotation period.

\subsection{Unsaturated M\,dwarfs on the extended slow rotator sequence}
\label{sec:newP}
\label{sec:Mdmem}

We now turn to an unexpected feature in the CAD -- the unsaturated early M\,dwarfs.
Among the stars in that region only one rotator was found in the original analysis of the light curves. 
We define this area as $(V-K_s)_0>3.7$ and $\log R'_\mathrm{IRT} < -4.05$. 
Using the  period estimation method described above, we find nine stars with periodic light curve variations. 
We examine these stars particularly closely because some of these rotation periods are longer than what we would normally have assigned with a 42\,d long time series. 
(Generally, we have required that two complete phases be visible over the observational baseline.) 
However, the stellar activity provides additional information that can justify these long periods.
In Table~\ref{tab:NewRot}, we list all such stars with activity informed rotation periods that can be found in the CAD on the declining branch among the M\,dwarfs.

\begin{table*} 
        \caption{Rotation periods for unsaturated M\,dwarfs and cluster membership criteria.}
        \label{tab:NewRot}
        \begin{tabular}{llrrrrrrl}
                \hline
                \hline
                ID & C11 & $(V-K_s)_0$ & $P_\mathrm{rot}$& $\Delta P_\mathrm{rot}$ & $p_\mathrm{RV}$ & $d$ & $d_\mathrm{pm}$ & Comment\\
                & & (mag) & (d) & (d) & & (pc) & (mas\,yr$^{-1}$)&\\
                \hline
                102025 & 248532 & 3.868 & 14.2 & 1.9 & 0.94 & 509 & 0.58 &\\
                304320 & 145504 & 3.938 & 32.0 & 3.5 & 0.86 & 497 & 0.23 & likely astrometric binary\\
                400454 & 179867 & 4.110 & 22.0 & 6.2 & 0.94 & 468 & 0.60 &\\
                400856 & 172827 & 3.718 & 20.6 & 1.8 & 0.66 & 390 & 0.97 & likely non-member\\
                522220 & 143090 & 4.295 & 30.0 & 3.5 & 0.94 & 521 & 1.02 & likely non-member\\
                604445 & 87263 & 3.889 & 17.1 & 1.6 & 0.93 & 469 & 0.30 &\\
                612827 & 48244 & 4.322 & 48.0 & 6.0 & 0.92 & 561 & 0.38 & photometric binary, likely non-member\\
                716466 & 184783 & 3.941 & 16.8 & 2.9 & 0.94 & 481 & 0.74 &\\
                801059 & 305389 & 4.187 & 27.0 & 3.6 & 0.93 & 471 & 0.17 &\\
                \hline 
        \end{tabular}
        \tablefoot{\emph{C11}: ID from \cite{2011AJ....141..115C}; $p_\mathrm{RV}$: Membership probability from radial velocity (F19); $d$ Distance from \cite{2018AJ....156...58B}, mean cluster distance $d=484$\,pc; $d_\mathrm{pm}$: distance from cluster centre in proper motion space $\mu_\alpha=-10.37$, $\mu_\delta=5.18$, based on \emph{Gaia~DR2} \citep{2018A&A...616A...1G}.}
\end{table*}

Despite the good agreement of the new slowly rotating stars in the general activity and the rotational properties of stars in NGC\,3532, we reinvestigated their membership status for additional assurance.
We applied stronger quality cuts than in our previous membership work (F19) in order to include only high-probability cluster members.

After reconsidering all membership criteria (Table~\ref{tab:NewRot}), we 
reclassified 
three stars as likely non-members because their distances 
locate them farther from the cluster centre than the others. (We still list them in Table~\ref{tab:NewRot} so that they can be reinstated if or when more definitive membership information becomes available.)
The remaining six stars are cluster members with very high probability. 
We note, for completeness, that all considered stars are photometric cluster members. 
However, the non-member 612827 is on the photometric binary sequence and the member 304320 is a potential astrometric binary ($\mathrm{RUWE}=1.64$).

We show the light curves for the six likely members, 
together with the generalized Lomb-Scargle \citep{2009A&A...496..577Z}, as well as the \textsc{clean} \citep{1987AJ.....93..968R, 2001SoPh..203..381C} periodograms in Fig.~\ref{fig:newLC}. The chosen periods do not always coincide with the highest peak in the periodograms for two reasons. Firstly, for the long rotation periods the peaks chosen are near or beyond the typical limit we set for our periods in the companion paper, where we require two cycles in our time series. Secondly, the rotation periods are assigned in concert with the chromospheric activity, which restricts the possible periods. In particular the two peaks that are seen in most periodograms in Fig.~\ref{fig:newLC} are typically aliases of each other. The chromospheric activity helps to break the half and double period alias effectively.

\subsection{New periods in colour-activity-rotation space}

\begin{figure*}
        \includegraphics[width=\textwidth]{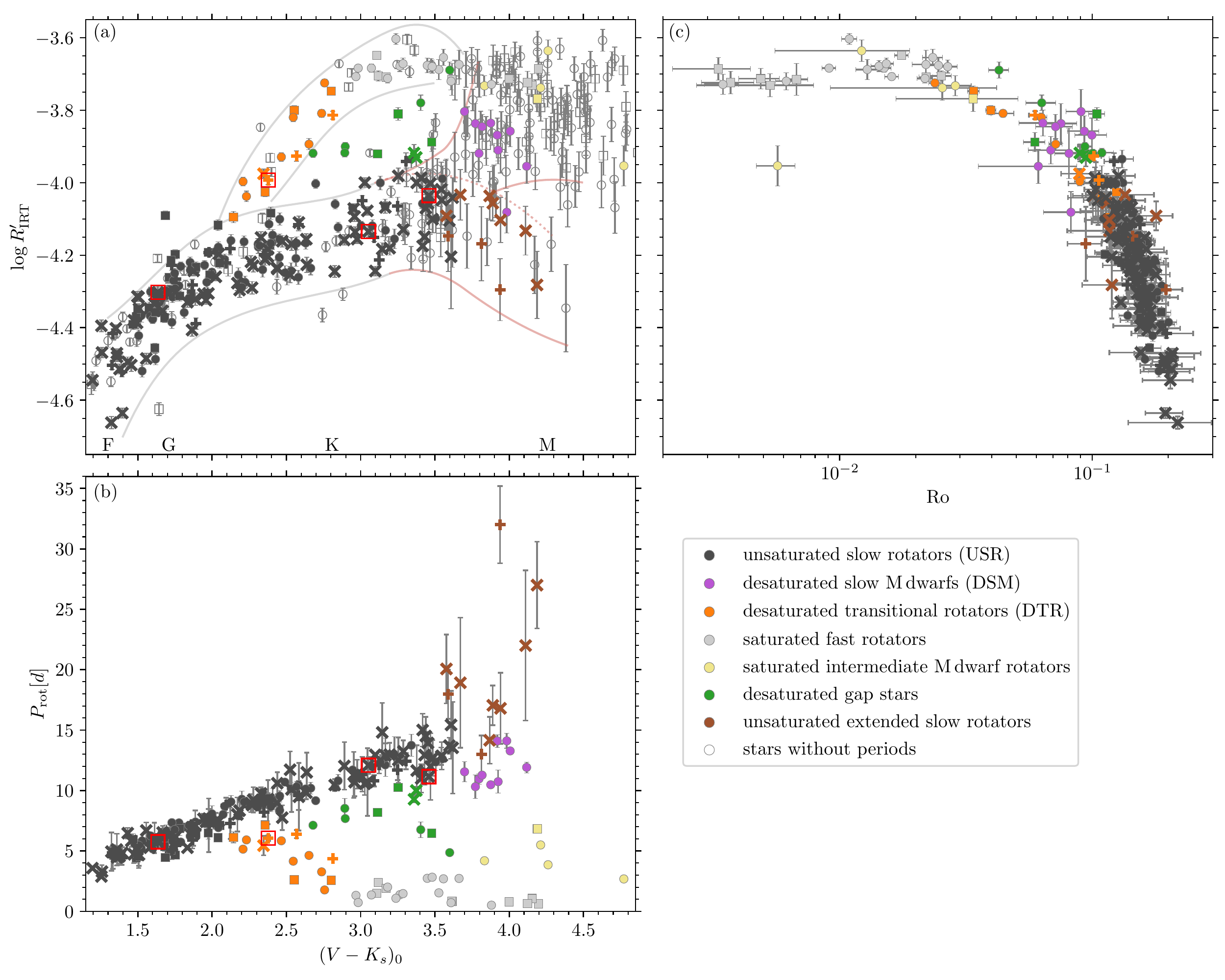}
        \caption{Same as Fig.~\ref{fig:IRTactivity} but now including the newly identified (activity-informed) rotation periods; single stars are marked with crosses and possible binaries with plus symbols. Each new period is also colour-coded to match the previously determined rotational groups, enabling cross-identification as in Fig.~\ref{fig:IRTactivity}. 
    This diagram includes the unsaturated  extended slow rotator sequence (brown), absent in Fig.~\ref{fig:IRTactivity}. The light curves and periodograms of the four stars marked with red squares (both in the CAD and CPD) are displayed in Fig.~\ref{fig:exampleLC} of the appendix to demonstrate the viability of deriving these activity-informed periods.}
        \label{fig:newIRTactivity}
\end{figure*}

With the large number (\nperact{}) of new rotation periods, we are able to update the central figure of this work (Fig.~\ref{fig:IRTactivity}), and show the complete picture, including the slowly rotating M\,dwarfs, in Fig.~\ref{fig:newIRTactivity}.
We note that the \emph{Gaia} parallaxes for 13 stars place them far from the cluster centre, despite all other membership criteria being fulfilled. 
We have conservatively excluded these stars from the rotational analysis despite their being potential cluster members, so as to avoid any contamination of our sample.

The newly found periods can be observed to strengthen all key regions of the CAD with the exception of the saturated stars.
Most notably, we are now able to assign periods to the least active, highest-mass stars of our sample, thus finding many additional periods along the whole slow rotator sequence, to fill it out substantially. 
In the corresponding rotation-activity diagram (Fig.~\ref{fig:newIRTactivity}c) all new stars can be seen to follow the same relation as the previously identified rotators. 
Since the fit to this diagram was the constraint on the new periods, this result is not unexpected, and serves mainly as a check on the results. 

The CPD is where we find the most noticeable changes. 
Not only is the slow rotator sequence much more densely populated than before, but F-type stars are also now present, and the relationship now also extends into the M\,dwarf regime out to periods of $\sim$30\,d. Slowly rotating M\,dwarfs in this period regime were also discovered in the Pleiades \citep{2016AJ....152..113R} and NGC\,2516 (F20). The stars in NGC\,3532 have certainly spun down, and we analyse their rotational evolution in the companion paper (F21rot).
The fact that these stars are both located on the rotation-activity relation and extend the slow rotator sequence seamlessly likely indicates that they share a dynamo-regime with the earlier-type stars despite being still on the pre-main sequence.
A similar conclusion for main sequence stars was also drawn by \cite{2016Natur.535..526W}, working with a sample of older, slowly rotating field M\,dwarfs. An in-depth analysis of the whole CPD is provided in the companion rotational study. 

%
%

\section{Conclusions}
\label{sec:conclusion}

We have measured the chromospheric activity of cool stars in the 300\,Myr old, rich open cluster NGC\,3532 using the \ion{Ca}{II} infrared triplet lines and have used these to explore and better define the corresponding colour-activity-rotation relationships.
The cluster occupies an important transitional location between the well-studied ZAMS clusters (e.g. Pleiades, NGC\,2516) and the 600\,Myr old Hyades cluster, and its richness makes it especially valuable.
This study follows on the heels of our prior spectroscopic membership study (F19) and a detailed study of the rotation period distribution in the companion paper (F21rot).

We showed how the chromospheric activity is well correlated with the photospheric activity, the latter measured from the starspot-induced variability amplitude of light curves. 
The correlation holds despite the fact that the two datasets are not contemporaneous, an encouraging sign from the observational perspective, and is ultimately rooted in their shared stellar rotation and its mass and age dependences.

We constructed one of the richest 
colour-activity diagrams (CAD) available to date for any open cluster.
The diagram is remarkably rich in information, with a prominent (main) sequence of lower-activity stars and a side branch of higher-activity stars, the two separated by an `activity gap'.
The CAD appears to be a near-mirror image of the colour-period diagram (CPD) of the cluster.
Colour-coding specific groups of rotators to show the correspondence emphasises that such a correspondence occurs on a star-by-star, rather than statistical, basis.
This correspondence suggests that the two quantities (rotation and activity) are potentially predictable from each other to a greater extent than previously thought possible.

In particular, the close correspondence between the activity and the rotation period shows that the absence of higher-mass fast rotators in our photometric work is an astrophysical fact and not a detection issue. 
We conclude that all stars down to a certain mass seem to have transitioned to the slow rotator sequence, showing the universality and strong mass dependence of this transition.
The CAD hosts an unexpected group of low-activity M\,dwarfs, for which the stellar activity declines with declining mass. 
In contrast to the majority of young M\,dwarfs, these stars are clearly in the unsaturated regime.

We also constructed the first rotation-activity diagram for stars in NGC\,3532 and, as expected, find the two well-known regimes (saturated and unsaturated) connected by a very wide transition region that can be traced back to the rotational gap in the CPD.
The relationship is remarkably tight in relation to prior work, and there is only one outlier.

From this rotation-activity correlation, we predicted rotation periods for stars in the unsaturated regime, searched for and identified the photometric rotation periods using the photometric light curves obtained for the companion rotational study. 
We recovered all previously found periods and were able to find \nperact{} additional membership-justified rotation periods 
with the help of this activity-informed period search. 
Using the activity as a guideline allows us to break aliases and to assign photometric rotation periods to noisier light curves. 
The new periods include stars on the extended slow rotator sequence.

In conclusion, the analysis of chromospheric activity in combination with photometric time series analysis enabled us to perform a detailed study of 
the activity-rotation period correspondence over a wide range of stellar masses among cool stars in NGC\,3532,
showing that the relationship is tight enough to be predictive over a wide parameter range. 
In particular, it allows us to substantially increase the number of photometric rotation periods derived. 
In combination with prior work, this study has contributed towards the establishment of NGC\,3532 as a benchmark open cluster and has wider implications for the era of massive multi-object spectroscopic surveys.

\begin{acknowledgements}
We thank Marcel Ag\"ueros for the detailed review.
SAB acknowledges support from the Deutsche Forschungs Gemeinschaft (DFG) through project number STR645/7-1. SPJ is supported by the German Leibniz-Gemeinschaft, project number P67-2018.
Based on data acquired at the Anglo-Australian Telescope under program S/2017A/02. We acknowledge the traditional owners of the land on which the AAT stands, the Gamilaraay people, and pay our respects to elders past and present.
Based in part on observations at Cerro Tololo Inter-American Observatory, National Optical Astronomy Observatory (2008A-0476; S.~A.~Barnes, SMARTS consortium through Vanderbilt University), which is operated by the Association of Universities for Research in Astronomy (AURA) under a cooperative agreement with the National Science Foundation.
This work has made use of data from the European Space Agency (ESA) mission
\emph{Gaia} (\url{https://www.cosmos.esa.int/gaia}), processed by the \emph{Gaia}
Data Processing and Analysis Consortium (DPAC,
\url{https://www.cosmos.esa.int/web/gaia/dpac/consortium}). Funding for the DPAC
has been provided by national institutions, in particular the institutions
participating in the \emph{Gaia} Multilateral Agreement.
This research has made use of NASA's Astrophysics Data System Bibliographic Services.
This research has made use of the SIMBAD database and the VizieR catalogue access tool, operated at CDS, Strasbourg, France.
\textbf{Software:}
This research made use of \textsc{Astropy}, a community-developed core Python package for Astronomy \citep{2013A&A...558A..33A}.
This research made use of the following \textsc{Python} packages:
\textsc{Pandas} \citep{pandas};
\textsc{NumPy} \citep{numpy};
\textsc{MatPlotLib} \citep{Hunter:2007};
\textsc{IPython}: \citep{ipython};
\end{acknowledgements}

\bibliographystyle{aa} 

\bibliography{N3532-activity} 

\begin{appendix}
        \section{Light curves of selected stars}

        \begin{figure*}
        \includegraphics[width=\textwidth]{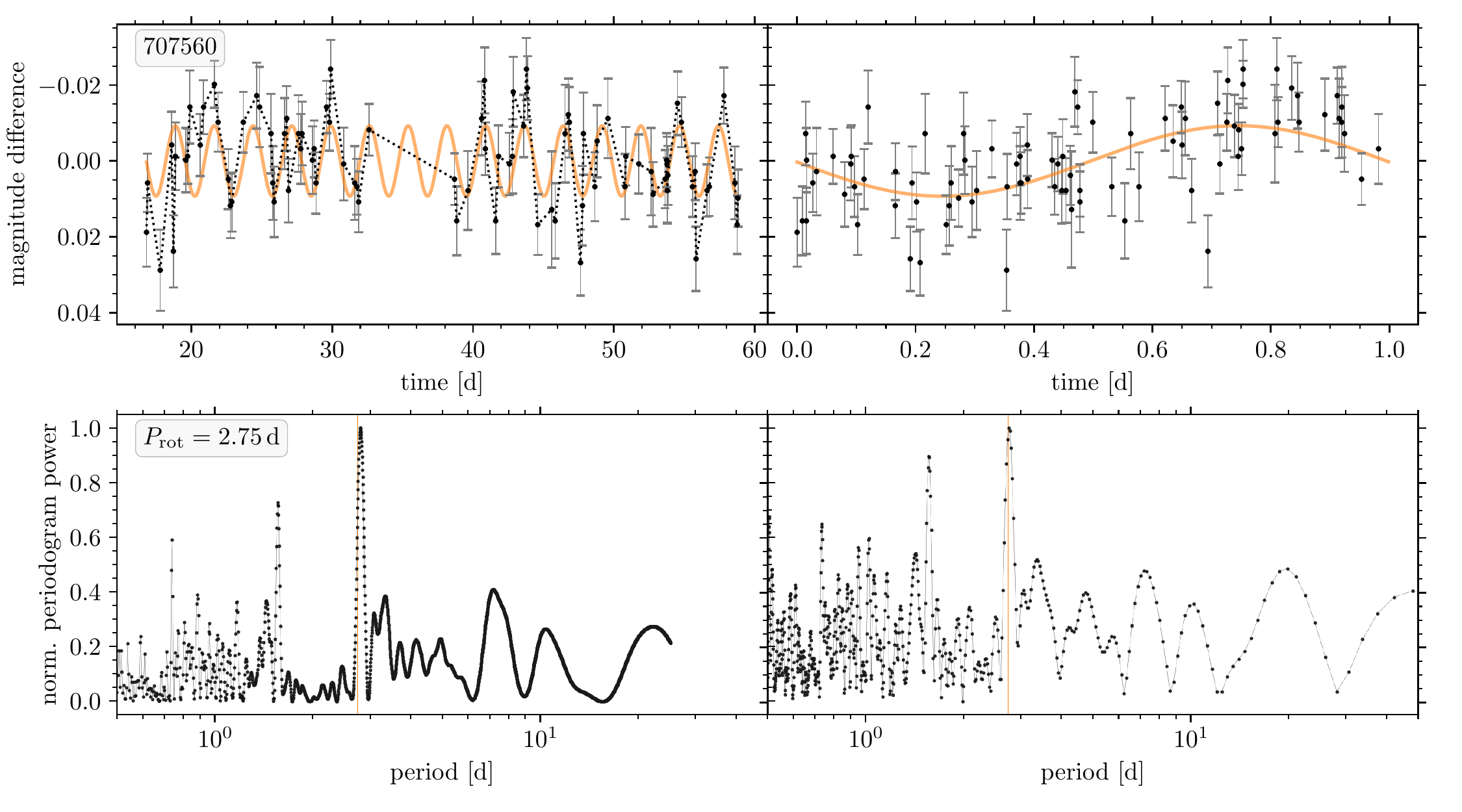}
        \caption{Light curve in time (\emph{top left}) and phase space (\emph{top right}) as well as the Lomb-Scargle periodogram (\emph{bottom left}) and \textsc{clean} periodogram (\emph{bottom right}) for the outlier (707560) in the rotation-activity diagram.}
        \label{fig:LCfullyconvective}
        \end{figure*}

        \begin{figure*}
        \includegraphics[width=\textwidth]{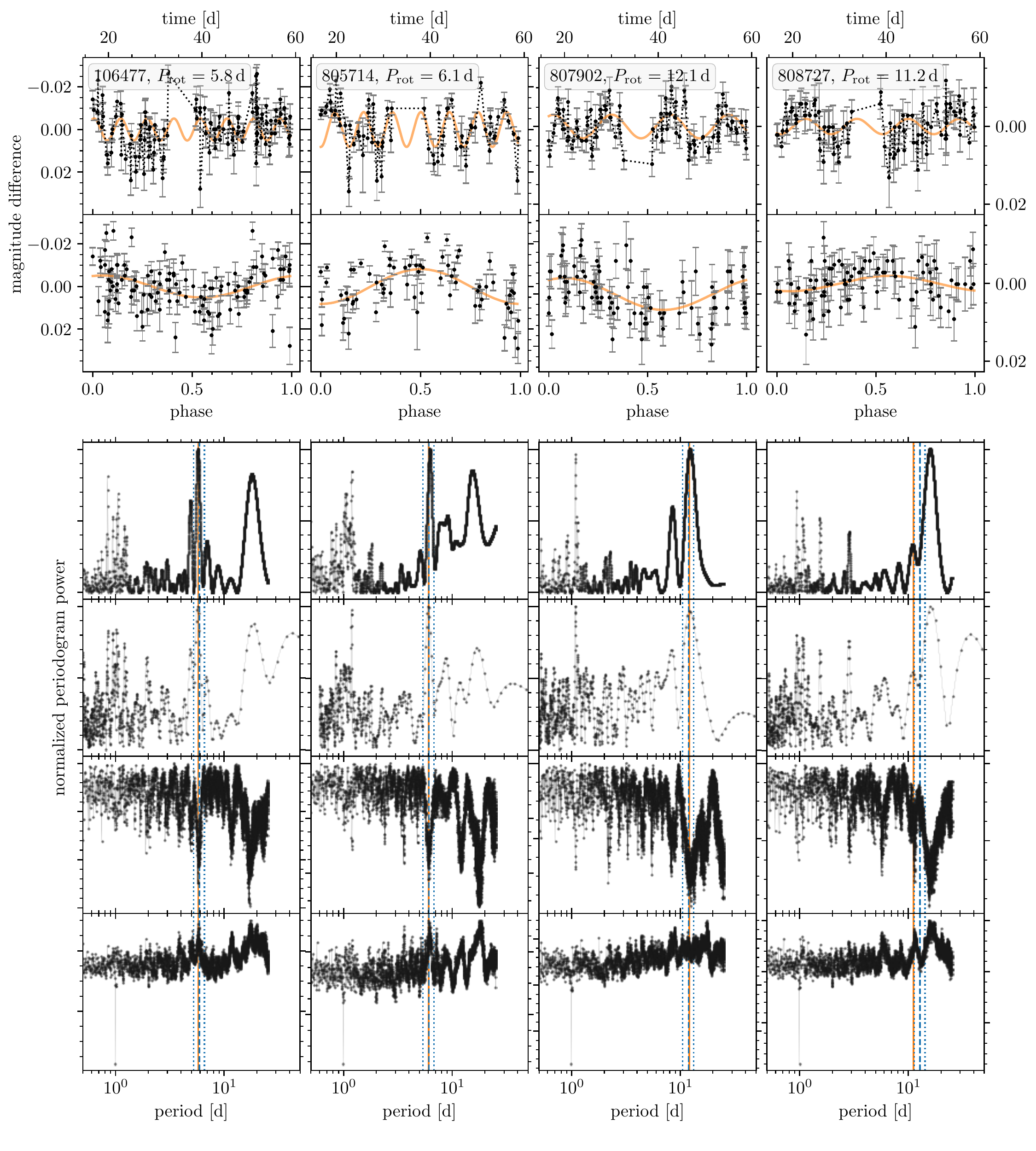}
        \caption{Four examples of light curves and their associated periodograms for (non-M) stars with newly identified, activity-informed rotation periods. Each column shows the data for one star. The two \emph{upper} panels show the light curves in both the time and phase domains. We fit a sine with the rotation period and display it along with the data. The left scale is valid for the two leftmost plots, while the right scale is for the two light curves on the right. The \emph{lower} panels show, from top to bottom, the Lomb-Scargle, \textsc{clean}, phase-dispersion minimization, and Gregory-Loredo periodograms. In each periodogram, we mark both the rotation period (orange line) and the period estimated from the activity and its uncertainty (blue lines). Unlike the other periodograms, in the phase-dispersion minimization the best periods are represented by dips.}
        \label{fig:exampleLC}
        \end{figure*}

        \begin{figure*}
        \includegraphics[width=\textwidth]{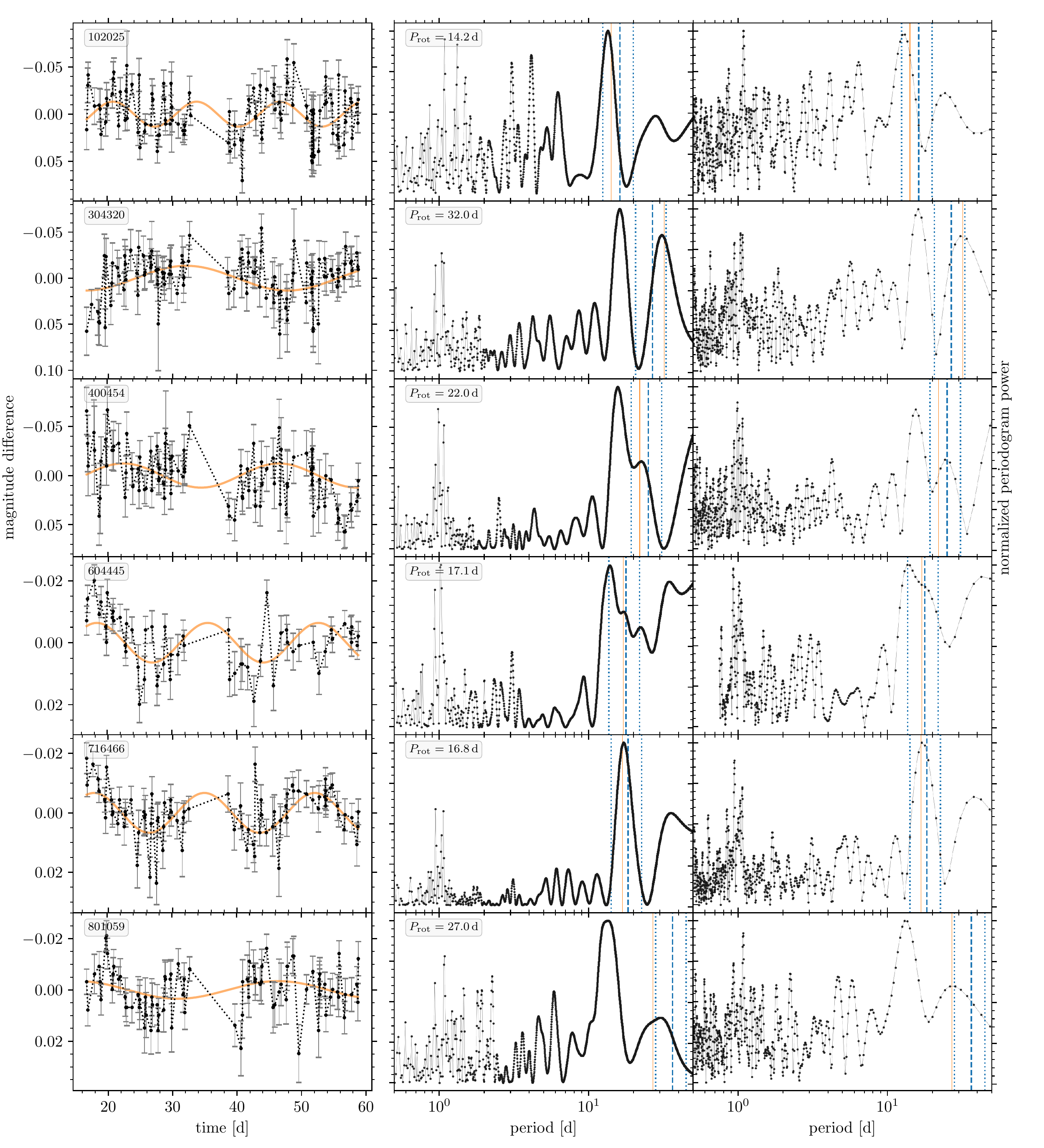}
        \caption{\emph{Left column}: Light curves for the early M\,dwarfs of low stellar activity in NGC\,3532 that have newly identified rotation periods. The orange line is a sine fit to the data with the newly identified likely rotation period. The first two light curves (from the top) are from $V$ 120\,s exposures, while all others are from $I_c$ 600\,s images.
        \emph{Centre column}: Corresponding normalised generalized Lomb-Scargle diagram, with the selected rotation period marked by the orange vertical line. The dashed blue line indicates the rotation period estimated from the chromospheric activity measurement, and the dotted lines delineate the search range, which is derived from the period uncertainty.
        \emph{Right column}: Same as the centre column but with the \textsc{clean} periodogram.
        The offset between the peaks and the position of the best period for 102025 in the first row is due to the different peak positions in the phase-dispersion minimization and Gregory-Loredo periodograms (not displayed).}
        \label{fig:newLC}
        \end{figure*}

\end{appendix}

\end{document}